\documentclass[twocolumn, aps, prd,   
               preprintnumbers,
               numbers,
               sort&compress,
               nofootinbib,
               showpacs,
               showkeys,
               colorlinks,
               linkcolor=blue,
               citecolor=blue]{revtex4-1} 

\usepackage{amssymb,graphicx}
\usepackage{amsmath}
\usepackage[]{empheq}
\usepackage{hyperref}
\usepackage{natbib,ifthen}

\usepackage{wasysym}
\usepackage{mathrsfs}
\usepackage[lofdepth,lotdepth,caption=false]{subfig}
\usepackage{color}
\usepackage{bbold}

\newcommand{\exclude}[1]{}

\newcommand{\commentOut}[1]{}

\newcommand{\be}{\begin{equation}}
\newcommand{\ee}{\end{equation}}
\newcommand{\beq}{\begin{eqnarray}}
\newcommand{\eeq}{\end{eqnarray}}

\newcommand{\jcap}{JCAP}

\begin{document}

\title{Mysterious anomalies in Earth's atmosphere and strongly interacting Dark Matter}

\author{Ariel  Zhitnitsky}
\email{arz@phas.ubc.ca}
\affiliation{Department of Physics and Astronomy, University of British Columbia, Vancouver, V6T 1Z1, BC, Canada}

\author{Marios Maroudas}
\affiliation{Institute of Experimental Physics, University of Hamburg, 22761 Hamburg, Germany}

\begin{abstract}
 
It has been recently argued in \cite{Bertolucci:2016xjm, Zioutas:2020ndf, Zioutas:2023ybw} that numerous enigmatic observations remain challenging to explain within the framework of conventional physics. These anomalies include unexpected correlations between temperature variations in the stratosphere, the total electron content of the Earth's atmosphere, and earthquake activity on one hand, and the positions of planets on the other. Decades of collected data provide statistically significant evidence for these observed correlations. The work in \cite{Bertolucci:2016xjm, Zioutas:2020ndf, Zioutas:2023ybw} suggests that these correlations arise from strongly interacting ``streaming invisible matter'' which gets gravitationally focused by the solar system bodies including the Earth's inner mass distribution. Here, we propose that some of these, as well as other anomalies, may be explained by rare yet energetic events involving the so-called axion quark nuggets (AQNs) impacting the Earth. In other words, we identify the ``streaming invisible matter'' conjectured in \cite{Bertolucci:2016xjm, Zioutas:2020ndf, Zioutas:2023ybw} with AQNs, offering a concrete microscopic mechanism to elucidate the observed correlations. It is important to note that the AQN model was originally developed to address the observed similarity between the dark matter and visible matter densities in the Universe \cite{Zhitnitsky:2002qa, Zhitnitsky:2021iwg}, i.e., $\Omega_{\rm DM} \sim \Omega_{\rm visible}$ and not explain the anomalies discussed here. Nonetheless, we support our proposal by demonstrating that the intensity and spectral characteristics of AQN-induced events are consistent with the aforementioned puzzling observations.
     
\end{abstract}  


\maketitle

\section{Introduction}
\label{sec:introduction}

The title of this work juxtaposes two seemingly contradictory concepts: ``strongly interacting" and ``dark matter" (DM). By definition, DM should decouple from baryons and radiation to play its cosmological role. Therefore, DM was considered as something that interacts feebly with normal matter and does not emit, absorb, or reflect light. However, the goal of this study is to reconcile this apparent contradiction by exploring the axion quark nugget (AQN) DM model \cite{Zhitnitsky:2002qa, Zhitnitsky:2021iwg}, which inherently behaves as a chameleon. AQNs are hypothetical very dense and microscopically large composite objects with a mass on the order of grams and sub-micrometer size. They consist of a core of quarks/antiquarks in a color-superconducting (CS) state surrounded by an electrosphere of positrons/electrons, and enclosed by an axion domain wall (see Fig.~\ref{AQN-structure} and Sect.~\ref{AQN} for more details). In diffuse environments, AQNs act as conventional DM candidates with negligible interactions. Yet, in dense environments, such as when they interact with planets or stars, they become strongly interacting objects capable of producing profound effects. 

A growing number of puzzling observations present anomalies that defy conventional physics interpretations. These anomalies include terrestrial phenomena like the unexpected seasonal stratospheric temperature variations, ionospheric anomalies, and their correlations with seismic activity, as well as solar phenomena like the coronal heating problem, the origin of sunspots, and the trigger mechanism of the solar flares. The additional observation of correlation patterns with the position of the solar system bodies further complicates their interpretation \cite{Bertolucci:2016xjm, Zioutas:2020ndf, Zioutas:2023ybw, Maroudas:2022ufl, Zioutas_2022_Radius}. Such phenomena challenge existing models of DM and atmospheric or planetary dynamics, leaving an explanatory gap. The aforementioned works propose the explanation of these anomalies through gravitational focusing by the solar system bodies including the inner Earth's mass distribution \cite{Hoffmann_2003_gravitational, Kryemadhi_2023_gravitational} of strongly interacting, invisible matter in the form of streams. This work identifies such ``invisible matter" as AQNs, providing a specific microscopic mechanism that aligns with these observations.

The AQN model offers an explanation for these anomalies by unifying them under a single framework. Unlike conventional DM candidates such as weakly interacting massive particles (WIMPs)—which remain undetected despite extensive searches—the AQN model introduces dense, composite objects made of strongly interacting quark matter. These objects satisfy cosmological constraints in diffuse environments while becoming strongly interacting in dense conditions, potentially explaining the energy deposition and correlations observed in Earth's atmosphere and ionosphere. Moreover, if future, more precise analyses (see e.g. proposals \cite{Cantatore:2020obc, Zioutas:2021xcm, Lazanu:2024ddm, Zioutas:2024cmy, Adair:2024wki}) confirm the puzzling observations discussed above, it could provide extraordinary evidence revealing the true nature of DM. This study demonstrates that the AQN framework can quantitatively account for the observed phenomena while aligning with previously proposed explanations based on gravitational focusing.

In summary, the implications of this study extend beyond resolving observational puzzles. If validated, the AQN model could address a deeper, fundamental question in modern cosmology—the nature of DM. By connecting astrophysical observations with terrestrial anomalies, this work aims to provide critical insights into this long-standing mystery.

\subsection{Observed mysteries}
\label{mysteries}

Before presenting our arguments, we first highlight the mysterious properties of the observations reported in \cite{Bertolucci:2016xjm, Zioutas:2020ndf, Zioutas:2023ybw, Maroudas:2022ufl}. These phenomena are challenging to interpret within the framework of conventional astrophysical models. The key puzzles are summarized as follows:

\begin{enumerate}

\item {\bf Stratospheric Temperature Puzzle}:
\label{item:first}
Seasonal variations in the stratospheric temperature typically follow a smooth pattern, with a minimum in December and a maximum in June, as measured at isobaric levels of 3 hPa, 2 hPa, and 1 hPa (corresponding to altitudes of approximately 38.5 km, 42.5 km, and 47.5 km, respectively) in the Northern Hemisphere ($42.5^\circ N, 13.5^\circ E$). However, an unexpected and striking temperature peak consistently appears around December–January, as observed over a 33-year period (1986–2018). This anomaly, depicted in Fig.~1 of \cite{Zioutas:2020ndf}, contradicts the well-known annual modulation pattern of the conventional DM flux, which for the Northern hemisphere is expected to peak in June and reach its minimum in December \cite{Freese:2012xd}.

\item {\bf Solar Non-Correlation Puzzle}:
\label{item:second}
Solar UV and EUV emissions are known to dominate atmospheric dynamics and influence temperature variations. To rule out direct solar irradiation as the cause of the temperature anomalies mentioned above, measurements of the $F10.7$ radio line ($\approx 2.8~\text{GHz}$), which is a proxy of solar activity, and solar EUV emissions were analyzed. These observations excluded solar activity as the primary source of the anomalies, as illustrated in Fig.~3 of \cite{Zioutas:2020ndf}.

\item {\bf Planetary Correlation Puzzle}:
\label{item:third}
Detailed analyses in \cite{Zioutas:2020ndf} demonstrated significant correlations between the stratospheric temperature fluctuations and planetary positions (see Figs.~3 to 8). Additionally, a systematic test was performed, as shown in Fig.~9 of \cite{Zioutas:2020ndf}, revealing that daily stratospheric temperature variations of approximately $1.2\%$ could accumulate to as much as $43\%$ when planetary positions were accounted for, with $69\%$ representing the theoretical maximum. This finding strongly indicates that the temperature variations are not random but coherent effects linked to planetary alignments.

\item {\bf TEC Puzzle}:
\label{item:fourth}
The total electron content (TEC) of the Earth's atmosphere, which measures ionization levels, also displays unexpected anomalies. Daily TEC measurements from 1995 to 2012 reveal a pronounced planetary dependence, with a seasonal difference between December and June as large as $20\%$. This deviation, shown in Fig.~12 of \cite{Bertolucci:2016xjm}, cannot be explained by the Earth-Sun distance or seasonal effects, as TEC values are averaged globally over the Earth's surface for both hemispheres.

\item {\bf TEC–Earthquake Correlation Puzzle}:
\label{item:fifth}
A strong correlation between global TEC anomalies and inner Earth's seismic activity of magnitude $M \geq 8$ has been observed, as shown in Fig.~7 of \cite{Zioutas:2023ybw, Zioutas_geoscience} during the period 1995-2012. This correlation is particularly puzzling because the primary source of TEC anomalies—UV and EUV photons—should, in principle, have no direct connection to seismic events originating deep within the Earth's interior.

\end{enumerate}

These correlations appear highly anomalous and difficult to reconcile with conventional physics. The conjecture proposed in \cite{Bertolucci:2016xjm, Zioutas:2020ndf, Zioutas:2023ybw, Maroudas:2022ufl} attributes these mysteries to ``streaming invisible matter," though without specifying its microscopic properties (e.g., mass, size, interaction patterns, or coupling constants).

\subsection{Normalization factors}
\label{noram_factors}

The analysis in \cite{Bertolucci:2016xjm, Zioutas:2020ndf, Zioutas:2023ybw, Maroudas:2022ufl} established two key benchmarks for understanding these phenomena. First of all, the energy required to explain the observed stratospheric temperature anomalies has been estimated as \cite{Zioutas:2020ndf}: 
\be
\label{E-deposition} 
\text{[Energy deposition]} \approx (0.1 - 1)~\frac{\text{W}}{\text{m}^2}. 
\ee

Secondly, the flux enhancement factor $A(t)$ for the ``streaming invisible matter," relative to the canonical DM flux in the Standard Halo Model (SHM) of $\sim 0.45\,\text{GeV}/\text{cm}^3$, is challenging to extract from observations. This difficulty arises from the complex interplay of several unknown factors, such as the velocity distribution of the ``streaming invisible matter" and its dependence on planetary positions including its actual interaction strength with normal matter. Additionally, the accumulation history of such matter over billions of years within the solar system could have significantly altered the structure and dynamics of these streams by the present time. Our normalization for the canonical SHM framework corresponds to $A=1$. From \cite{Zioutas:2021xcm} we quote below $A(t)\approx 10^4$ which corresponds to the flux enhancement by the Moon towards the Earth. However, it could locally and temporarily reach much larger values, ideally up to $10^8$ or even much more following combined gravitational focusing effects \cite{Hoffmann_2003_gravitational, Kryemadhi_2023_gravitational}.
\be
\label{enhancement} 
\text{[DM Flux Enhancement]}:~ A(t) \approx 10^4.
\ee

\subsection{The AQN hypothesis}
\label{aqn_hypothesis}

In this work, we propose a specific microscopic mechanism that fits-in as an explanation for all the profound and mysterious correlations discussed earlier, in items \ref{item:first}-\ref{item:fifth}, within a unified framework. This framework relies on the AQN model, with the same set of parameters fixed by observations in entirely different contexts. The ability of AQNs to produce such strong effects arises from their unique construction: these DM candidates are composed of strongly interacting quarks and gluons—identical to the constituents of conventional visible matter in quantum chromodynamics (QCD).

From a cosmological perspective, DM and ordinary matter exhibit a fundamental difference aside from their ``visibility". DM played a crucial role in the formation of the large-scale structure of the Universe. Without DM, the Universe would have remained too uniform to form galaxies and other cosmic structures. Ordinary matter alone could not produce sufficient fluctuations, as it remained tightly coupled to radiation until relatively recent epochs, preventing clustering. A key parameter governing DM's behavior in the Universe is the ratio of the interaction cross-section $\sigma$ to mass $M_{\rm DM}$, which must remain sufficiently small to satisfy cosmological constraints, see e.g. recent review \cite{Tulin:2017ara}:
\be
\label{sigma/m}
\frac{\sigma}{M_{\rm DM}}\ll  1\frac{\rm cm^2}{\rm g}
\ee

WIMPs satisfy this criterion due to their extremely small cross-section $\sigma$ for typical masses $M_{\rm WIMP} \in (10^2 - 10^3)~\rm GeV$. However, after dominating theoretical and experimental DM research for over four decades, the WIMP paradigm has failed. Despite significant improvements in detector sensitivity, no traces of WIMPs have been found, prompting the exploration of alternative DM models.

In this work, we consider a fundamentally different type of DM in the form of dense, macroscopically large composite objects known as AQNs. These objects, first introduced in \cite{Zhitnitsky:2002qa}, share similarities with Witten's quark nuggets \cite{Witten:1984rs, Farhi:1984qu, DeRujula:1984axn} but exhibit unique properties. AQNs behave as chameleons: in dilute environments, they are effectively non-interacting, satisfying the condition from Eq.~(\ref{sigma/m}) during the structure formation of the Universe, with: $\sigma /M_{\rm AQN}\lesssim 10^{-10} {\rm cm^2}{\rm g^{-1}}$ (see Eq.~(\ref{sigma/M})). In contrast, when AQNs encounter dense environments such as planets or stars, they interact strongly with ordinary matter, producing observable effects.

The AQN model was originally developed \cite{Zhitnitsky:2002qa} to address the observed similarity between the DM and visible matter densities in the Universe, $\Omega_{\rm DM} \sim \Omega_{\rm visible}$. Its motivation can be summarized as follows: it is commonly assumed that the Universe began with a symmetric state of zero net baryonic charge, which evolved into a baryon-asymmetric state via a baryogenesis process involving baryon number violation, non-equilibrium dynamics, and $\cal{CP}$ violation. In contrast, the AQN framework posits that baryogenesis is instead a charge segregation process rather than charge generation, where the global baryon number of the Universe remains zero at all times.

A key distinction between the AQN model and Witten's quark nuggets \cite{Witten:1984rs, Farhi:1984qu, DeRujula:1984axn} is that AQNs can consist of both {\it matter} and {\it antimatter}, formed during the QCD phase transition as a result of this charge segregation process (see Sect.~\ref{basics} for more details). This {\it antimatter} component implies the existence of rare but profound annihilation events when antimatter AQNs collide with ordinary matter (see estimates in Sect.~\ref{AQN}). The claim of the present work is that these events could naturally account for the energy deposition from Eq.~(\ref{E-deposition}) that is required to explain, for example, the stratospheric temperature anomalies observed in \cite{Zioutas:2020ndf}.

The anomalous observations listed in items \ref{item:first}-\ref{item:fifth} are not the only mysteries that may be associated with DM streams. Other unusual and poorly understood phenomena, such as cosmic-ray-like events, might also be explained within the AQN framework. For example:

\begin{itemize}
    \item The Telescope Array (TA) collaboration \cite{Abbasi:2017rvx, Okuda_2019} reported mysterious bursts where multiple air showers occurred within 1 ms, defying explanation by conventional high-energy cosmic rays (CR). These bursts can be interpreted as AQN-induced events \cite{Zhitnitsky:2020shd}.

    \item Similar ``exotic events" recorded by the AUGER collaboration \cite{PierreAuger:2021int, 2019EPJWC.19703003C, Colalillo:2017uC} are also consistent with the AQN hypothesis \cite{Zhitnitsky:2022swb}.

    \item The ANITA experiment detected two anomalous upward-propagating events with non-inverted polarity \cite{Gorham:2016zah, Gorham:2018ydl}, which can also be explained via the AQN framework \cite{Liang:2021rnv}.

    \item The list of anomalous atmospheric events also includes the  Multi-Modal Clustering  Events observed by HORIZON 10T \cite{2017EPJWC.14514001B, Beznosko:2019cI}. These are very hard to understand in terms of the conventional CR events but could be interpreted in terms of the  AQN annihilation events in the atmosphere as argued in \cite{Zhitnitsky:2021qhj}. 
\end{itemize}
Additionally, AQNs may generate acoustic and seismic signals \cite{Budker:2020mqk} that could be in principle recorded if several dedicated instruments are present in the area. For example, a powerful seismic and atmospheric event (a so-called ``sky-quake") was recorded in the infrasound frequency band by the Elginfield Infrasound Array (ELFO) and attributed to an AQN-induced phenomenon \cite{Budker:2020mqk}. While exceptionally powerful events like this are rare, they underscore the potential for AQNs to explain a wide range of terrestrial and atmospheric anomalies.

This study focuses specifically on the anomalies listed in items \ref{item:first}-\ref{item:fifth}, as these exhibit clear correlations with planetary positions and represent decades of data collected across the Earth's surface. In contrast, other anomalies, such as cosmic-ray-like events, are less statistically significant due to limited detection areas and event frequencies, making planetary correlations more challenging to identify.

This paper is organized as follows. In Sect.~\ref{AQN}, we provide a brief overview of the AQN framework, emphasizing the key elements that are most relevant to the present study. In Sect.~\ref{proposal}, we outline our hypothesis, identifying the ``streaming invisible matter" proposed in \cite{Bertolucci:2016xjm, Zioutas:2020ndf, Zioutas:2023ybw, Maroudas:2022ufl} as DM particles in the form of AQNs. Sect.~\ref{sect:consistency} examines our proposal in the context of the mysterious and puzzling observations summarized in items \ref{item:first}-\ref{item:fifth} above. We demonstrate how these observations can be naturally explained within the AQN framework, presenting an explicit microscopic model capable of accounting for the reported correlations from \cite{Bertolucci:2016xjm, Zioutas:2020ndf, Zioutas:2023ybw, Maroudas:2022ufl}. Finally, in Sect.~\ref{conclusion}, we summarize our findings and propose specific tests that could validate or refute our hypothesis. Additionally, we highlight other intriguing, unexplained phenomena at various scales—from the early Universe to the Solar System—that may also be connected to DM in the form of AQNs.

\section{The AQN DM Model}
\label{AQN}

In this section, we provide an overview of the AQN DM model. Subsection \ref{basics} introduces the fundamental principles underlying the AQN framework, while Subsection \ref{AQN-dense} highlights specific features of AQNs relevant to the present study.

\subsection{The Basics}
\label{basics}

As mentioned earlier, the AQN construction shares similarities with Witten's quark nugget model \cite{Witten:1984rs,Farhi:1984qu,DeRujula:1984axn}. Both models propose DM candidates that are ``cosmologically dark" due to the small ratio in Eq.~(\ref{sigma/m}). This small ratio suppresses observable consequences of an otherwise strongly-interacting DM candidate in the form of the AQN nuggets.  

Indeed, for typical AQN parameters, the relevant ratio  assumes the following numerical value \cite{Zhitnitsky:2021iwg}:
\be \label{sigma/M} \frac{\sigma}{M_N} \sim \frac{\pi R^2}{M_N} \sim 10^{-10}{\rm cm^2g^{-1}},
\ee
where $R$ is the nugget's size, and $M_N$ is its mass. For this estimate, we use the typical parameters for AQNs provided in Tab.~\ref{tab:basics}. This value satisfies the cosmological constraint from Eq.~(\ref{sigma/m}) while allowing AQN to behave as a viable DM candidate.

A critical feature of AQNs is their flux, which determines the frequency of interactions and underpins the observable effects discussed earlier (items \ref{item:first}-\ref{item:fifth}). The flux can be estimated as \cite{Lawson:2019cvy}:
 \be
\label{Phi1}
\frac{\rm d \Phi}{\rm d A}
=\frac{\Phi}{4\pi R_\oplus^2}  =  4\cdot 10^{-2}\left(\frac{10^{25}}{\langle B\rangle}\right)\rm \frac{events}{yr\cdot  km^2},
\ee
where $R_\oplus = 6371~\rm km$ is the Earth's radius, $\langle B \rangle$ is the average baryon charge of AQNs (discussed below), and $\Phi$ is the total hit rate of AQNs on Earth \cite{Lawson:2019cvy}:
\be
\label{Phi}
\Phi
\approx \frac{2\cdot 10^7}{\rm yr}  
 \left(\frac{\rho_{\rm DM}}{0.3{\rm\,GeV\,cm^{-3}}}\right)
\left(\frac{v_{\rm AQN}}{220~ \rm km ~s^{-1}}\right)
\left(\frac{10^{25}}{\langle B\rangle}\right), \nonumber
\ee
where $\rho_{\rm DM}$ is the local DM density in the SHM, and $v_{\rm AQN}$ is the AQN velocity. In the ``invisible stream" hypothesis, the rate (\ref{Phi1}) has to be multiplied with the factor $A(t)$, as defined in Eq.~(\ref{enhancement}). Similar flux estimates also apply to Witten's quark nuggets \cite{Witten:1984rs, Farhi:1984qu, DeRujula:1984axn}.

Despite these similarities, the AQN model introduces several key elements that address the limitations of earlier constructions \cite{Witten:1984rs, Farhi:1984qu, DeRujula:1984axn}:
\begin{itemize}
    \item {\it Matter and antimatter composition}: Unlike the original quark nugget model, AQNs can consist of both {\it matter} and {\it antimatter}, formed during the QCD phase transition. This property enables rare but significant annihilation events when antimatter AQNs interact with ordinary matter, producing observable signatures.
    
    \item {\it Axion domain walls}: The AQN model incorporates axion domain walls formed during the QCD transition. These domain walls act as stabilizing structures, alleviating the need for a first-order phase transition as in the original Witten's model. The axion domain wall effectively ``squeezes" the nugget, providing additional stabilization absent in earlier models.
    
    \item {\it Vacuum energy}: In AQNs, the vacuum ground state energies inside the nugget (CS phase) differ significantly from outside the nugget (hadronic phase). This disparity enables the coexistence of these two phases only under external pressure, provided by the axion domain wall. This mechanism prevents nugget evaporation on the Hubble time scale, which was a key issue in Witten's original model \cite{Witten:1984rs, Farhi:1984qu, DeRujula:1984axn} which was assumed to be stable at zero external pressure.

    \item {\it Energy transfer}: Another pivotal difference between the AQN framework and Witten's model lies in the energy transfer to the surrounding material. For Witten's nuggets, the maximum energy transfer is constrained by the relatively low DM velocity $\sim 10^{-3}c$, limiting the energy transfer to $m_p v^2 / 2 \sim 10^{-6} m_p c^2$ per baryon charge of the nugget. In the AQN model, the available energy due to matter-antimatter annihilation can be as high as $2 m_p c^2 \approx 2~\rm GeV$ per baryon charge. This stark contrast in energy transfer dramatically alters the search strategies for such DM candidates, making AQNs far more observable through their energetic interactions.

    \item {\it Cosmological density problem and baryon asymmetry}: The AQN model inherently addresses the cosmological density problem by linking the DM density $\Omega_{\rm DM}$ (represented by matter and antimatter nuggets) and the visible matter density $\Omega_{\rm visible}$. Both densities are proportional to the same fundamental dimensional parameter of the theory, $\Lambda_{\rm QCD}$. Consequently, the model naturally predicts $\Omega_{\rm DM} \sim \Omega_{\rm visible}$ without requiring fine-tuning or additional parameters. By construction, the AQN framework resolves two fundamental problems in cosmology: the nature of DM and the baryon asymmetry of the Universe. The formation of AQNs, the generation of baryon asymmetry, and the survival pattern of the nuggets through the hostile environment of the early Universe have been extensively studied in \cite{Liang:2016tqc, Ge:2017ttc, Ge:2017idw, Ge:2019voa}. These works provide detailed insights into how AQNs form during the QCD phase transition, remain stable, and contribute to the observed cosmological densities.
\end{itemize}
By addressing these limitations and incorporating unique features, the AQN model not only provides a robust DM candidate, but also aligns naturally with fundamental cosmological observations.

\begin{table*}
\captionsetup{justification=raggedright}
	\begin{tabular}{cccrcc} 
		\hline\hline
		  Property  && \begin{tabular} {@{}c@{}}{ Typical value or feature}~~~~~\end{tabular} \\\hline
		  AQN's mass~  $[M_N]$ &&         $M_N\approx 16\,g\,(B/10^{25})$     \cite{Zhitnitsky:2021iwg}     \\
		   Baryon charge constraints~   $ [B]  $   &&        $ B \geq 3\cdot 10^{24}  $     \cite{Zhitnitsky:2021iwg}    \\
		   Annihilation cross section~  $[\sigma]$ &&     $\sigma\approx\kappa\pi R^2\simeq 1.5\cdot 10^{-9} {\rm cm^2} \cdot  \kappa (R/2.2\cdot 10^{-5}\rm cm)^2$  ~~~~     \\
		  Density of AQNs~ $[n_{\rm AQN}]$         &&          $n_{\rm AQN} \sim 0.3\cdot 10^{-25} {\rm cm^{-3}} (10^{25}/B) $   \cite{Zhitnitsky:2021iwg} \\
		  Survival pattern during BBN &&       $\Delta B/B\ll 1$  \cite{Zhitnitsky:2006vt,Flambaum:2018ohm,SinghSidhu:2020cxw,Santillan:2020lbj} \\
		  Survival pattern during CMB &&           $\Delta B/B\ll 1$ \cite{Zhitnitsky:2006vt,Lawson:2018qkc,SinghSidhu:2020cxw} \\
		  Survival pattern during post-recombination &&   $\Delta B/B\ll 1$ \cite{Ge:2019voa} \\\hline
	\end{tabular}
\caption{Basic properties of the AQNs adopted from \cite{Budker:2020mqk}. The parameter $\kappa$ is introduced to account for possible deviations from the geometric value $\pi R^2$ as a result of the ionization of the AQNs due to interaction with the environment. The ratio $\Delta B/B\ll 1$ implies that only a small portion $\Delta B$  of the total (anti)baryon charge $B$  hidden in form of the AQNs get annihilated during big-bang nucleosynthesis (BBN), Cosmic Microwave Background (CMB), or post-recombination epochs (including the galaxy and star formation), while the dominant portion of the baryon charge survives until the present time.}
	\label{tab:basics}
\end{table*}

The strongest direct detection limit\footnote{Non-detection of etching tracks in ancient mica provides an additional indirect constraint on the flux of DM nuggets with mass $M > 55$ g \cite{Jacobs:2014yca}. However, this constraint assumes a uniform nugget mass, which does not apply to the AQN model.} is set by the IceCube observatory, as detailed in Appendix A of \cite{Lawson:2019cvy}:
\be 
\label{direct}
\langle B \rangle > 3 \cdot 10^{24} ~~~[{\rm direct~(non)detection~constraint]}. 
\ee
This constraint is based on the non-detection of macroscopic DM nuggets by IceCube, with an effective surface area of $\sim \rm km^2$ during its 10 years of operation. The estimate in Eq.~(\ref{direct}) assumes a $100\%$ detection efficiency for macroscopically large nuggets, ruling out AQNs with baryon charges $\langle B \rangle < 3 \cdot 10^{24}$ at a confidence level of approximately $3.5\sigma$.

In \cite{SinghSidhu:2020cxw}, a broader set of constraints on antimatter nuggets was considered, without incorporating the specific features of the AQN model, such as the CS phase of the quark matter in the nugget core (see \cite{Alford:2007xm} for an overview of the CS phases). While the constraints in Eq.~(\ref{direct}) align with the findings from \cite{SinghSidhu:2020cxw}, including those based on CMB and BBN data, they differ regarding the so-called ``Human Detectors." As argued in \cite{Ge:2020xvf}, the estimates for ``Human Detectors" in \cite{SinghSidhu:2020cxw} are oversimplified and lack the robustness of constraints derived from CMB or BBN data.

\begin{figure}[h]
	\centering
	\captionsetup{justification=raggedright}
	\includegraphics[width=0.8\linewidth]{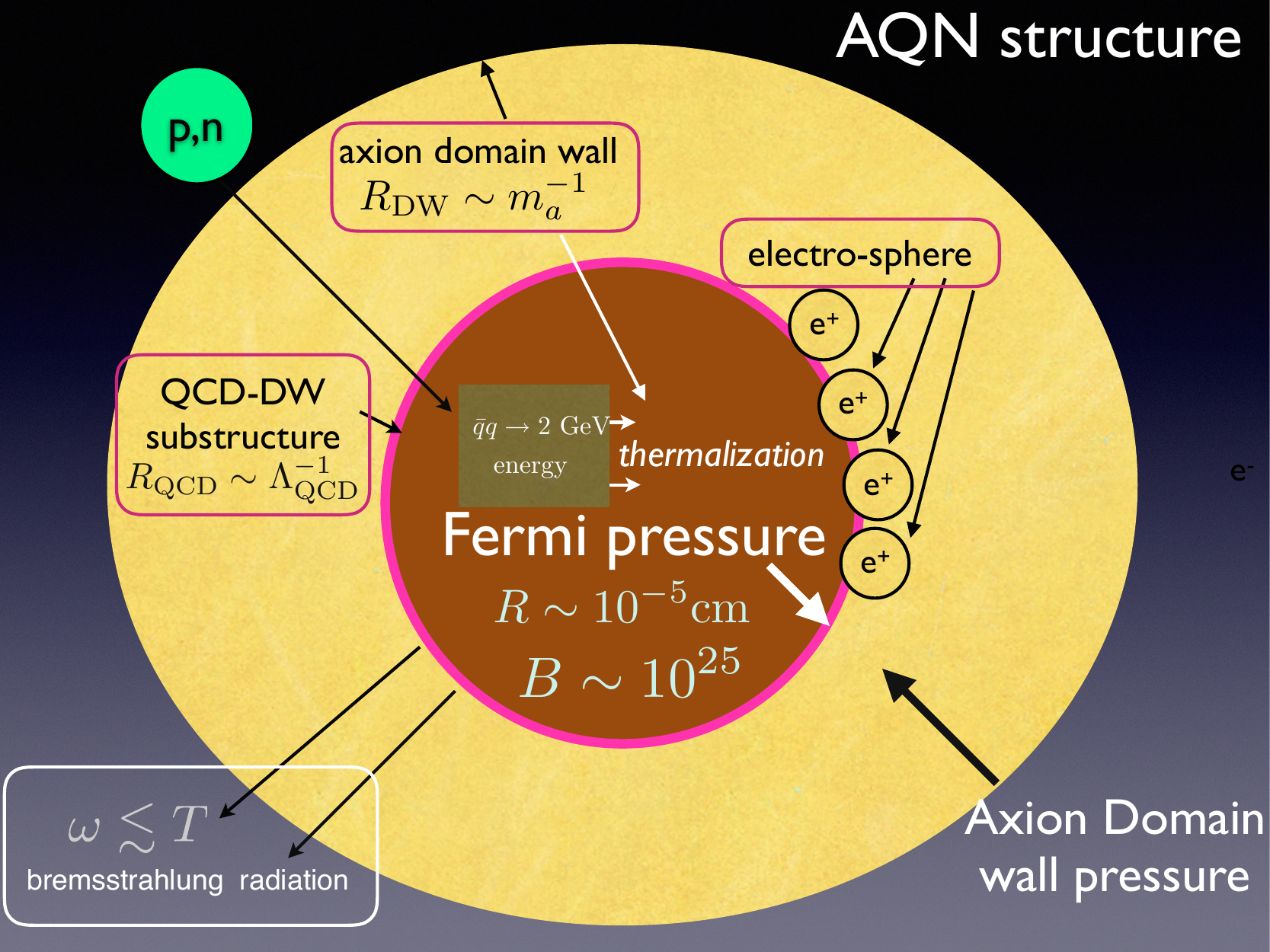}
	\caption{AQN structure (not in scale), adopted from \cite{Zhitnitsky:2022swb}. The dominant portion of the energy $\sim 2$ GeV produced as a result of a single annihilation process inside the anti-nugget is released in the form of the bremsstrahlung radiation with frequencies $\omega\leq T$. See the main text for detailed explanations.}
\label{AQN-structure}
\end{figure}

The internal structure of AQNs, depicted in Fig.~\ref{AQN-structure}, is critical to understanding their interactions and energy release mechanisms. Finally, Tab.~\ref{tab:basics} summarizes the basic properties and parameters of AQNs. It is noted that a key feature of the AQN model is that only a small fraction, $\Delta B \ll B$, of the total (anti)baryon charge $B$ hidden in the form of AQNs is annihilated over the Universe's long evolution, while the dominant portion of the baryon charge survives until the present time.


\subsection{When the AQNs hit the Earth}
\label{AQN-dense}

In the context of the present work, the most relevant studies concern the effects that may occur when AQNs composed of antimatter propagate through environments with sufficiently high visible matter density $n(r)$, such as the Earth's atmosphere or its deep interior. In these regions, annihilation processes are triggered, releasing a significant amount of energy into the surrounding material, which can manifest in various observable ways.

The interaction of AQNs with visible matter was initially studied in \cite{Forbes:2008uf} in the context of a galactic environment with a typical baryon density of $n_{\rm galaxy} \sim \mathrm{cm^{-3}}$. Here, we extend these computations to account for the higher-density environments of the Earth's atmosphere and interior, incorporating additional elements relevant to these cases.

When an AQN enters a region with baryon density $n$, annihilation processes begin, causing its internal temperature $T$ to rise. The internal temperature $T$ can be estimated from the condition that the radiative energy output balances the energy flux deposited onto the nugget:
\be
\label{eq:rad_balance}
    F_{\rm{tot}} (T) (4\pi R^2)
\approx \kappa\cdot  (\pi R^2) \cdot (2~ {\rm GeV})\cdot n \cdot v_{\rm AQN},  
\ee 
where $n$ is the baryon number density of the surrounding material, $F_{\rm{tot}}(T)$ is the total surface emissivity of the nugget (see Eq.~(\ref{eq:P_t})), $\kappa$ is a factor accounting for theoretical uncertainties in the annihilation processes of the (antimatter) AQN  colliding with surrounding material, and $v_{\rm AQN}$ is the velocity of the AQN. The left-hand side of Eq.~(\ref{eq:rad_balance}) represents the total radiative energy emitted per unit time from the nugget's surface while, the right-hand side accounts for the rate of annihilation events, with each successful annihilation event of a single baryon charge producing $\sim 2m_pc^2\approx 2~{\rm GeV}$ energy.

The total surface emissivity of an AQN due to bremsstrahlung radiation from its electrosphere at temperature $T$ was calculated in \cite{Forbes:2008uf} and is given by:
\begin{equation} 
\label{eq:P_t} 
    F_{\rm{tot}} \approx \frac{16}{3} \frac{T^4 \alpha^{5/2}}{\pi} \sqrt[4]{\frac{T}{m}},
\end{equation}
where $\alpha\approx1/137$ is the fine structure constant, $m = 511~\mathrm{keV}$ is the mass of electron, and $T$ is the internal temperature of the AQN. Notably, the radiation emitted from the electrosphere is non-thermal and differs significantly from blackbody radiation. The bremsstrahlung spectrum has unique characteristics that can help distinguish AQN interactions from other sources of energy release. It is noted that the thermal properties presented above were originally applied to the study of the emission from AQNs from the Galactic Centre, where a nugget's internal temperature is very low, $T\sim$ eV. 

When AQNs propagate through the Earth's atmosphere, their internal temperature begins to rise, reaching approximately $\sim 40~\rm keV$. As the AQNs penetrate the Earth's surface and travel toward its center, the much higher density of the Earth's interior causes the internal temperature to increase further, up to $\sim 400~\rm keV$. These temperature estimates can be derived using Eqs.~(\ref{eq:P_t}) and (\ref{eq:rad_balance}):
 \beq
 \label{T}
 T_{\rm atmosphere}&\approx& 40 ~{\rm keV} \cdot \left(\frac{n_{\rm atmosphere}}{ 10^{21} ~{\rm cm^{-3}}}\right)^{\frac{4}{17}}   {\kappa}^{\frac{4}{17}} \nonumber \\
  T_{\rm interior}&\approx& 400 ~{\rm keV} \cdot \left(\frac{n_{\rm interior}}{10^{25} ~{\rm cm^{-3}}}\right)^{\frac{4}{17}} {\kappa}^{\frac{4}{17}}.
 \eeq
For the purposes of this study, we adopt $T \approx 400~\rm keV$ as the benchmark temperature for AQNs exiting the Earth's surface in the upward direction.

Another key parameter for this analysis is the specific heat, $c_V$, which determines the total energy accumulated by an AQN during its passage through the Earth's interior. This energy is primarily a function of the exit temperature, $T$, and $c_V$. The specific heat depends on the CS phase of the quark matter inside the nugget, with different CS phases yielding distinct expressions for $c_V$. In the simplest two-flavor superconducting phase (2SC), the specific heat is given by \cite{Alford:2007xm}:
\be \label{2SC} c_V \simeq \frac{1}{3} T (\mu_d^2 + \mu_u^2),
\ee
where $\mu_u$ and $\mu_d$ are the chemical potentials for up and down quarks, respectively, in the CS phase. These chemical potentials are approximately $\mu_u \simeq \mu_d \simeq 500~\rm MeV$, consistent with earlier studies of the typical value of the AQN’s chemical potential at the moment \cite{Ge:2019voa}. For our numerical analysis, we will use Eq.~(\ref{2SC}) to estimate $c_V$.

The AQN framework, using the same set of parameters applied in this study, has the potential to explain numerous puzzling observations that remain unresolved by conventional astrophysical models. Many of these phenomena, observed on Earth, were highlighted at Sect.~\ref{mysteries}. Additionally, the AQN model could offer possible explanations for enigmatic observations across vastly different scales and cosmological epochs. These include events during the BBN epoch, the dark ages, as well as interactions in galactic and Solar environments. Further details and references can be found in the concluding Sect.~\ref{sect:paradigm}.

\section{Energy deposition in the Earth's atmosphere by AQNs}
\label{proposal}

Our objective is to propose that the mysterious correlations (items \ref{item:first}-\ref{item:fifth}) introduced earlier can be explained within the AQN framework. Specifically, we argue that these correlations arise from AQNs depositing enormous amounts of energy into the Earth's atmosphere during their propagation.

To frame this argument, we distinguish between two categories of (antimatter) AQNs contributing to these effects:
\begin{enumerate}
\item {\it Downward-moving AQNs}: These nuggets propagate from space into the Earth's atmosphere, reaching typical internal temperatures of $T_{\rm atmosphere} \approx 40~\rm keV$ as they interact with the atmospheric material.
\item {\it Upward-moving AQNs}: These nuggets traverse the Earth, exiting through the surface. Due to the higher density of the Earth's interior, their temperatures increase significantly, reaching $T_{\rm interior} \approx 400~\rm keV$ (as per Eq.~(\ref{T})).
\end{enumerate}
Our focus below is on the upward-moving AQNs with the highest possible temperatures, as these contribute the most energy.

Using the expression for specific heat, $c_V$, from Eq.~(\ref{2SC}), we estimate the total energy accumulated by a single AQN during its passage through the Earth's interior. Assuming $T \approx 400~\rm keV$ as the characteristic temperature, we find\footnote{\label{altitude}Most of this energy in Eg.~(\ref{AQN-accumulated}) is released into the surrounding atmospheric material almost immediately after the nugget exits the Earth’s surface. The rapid cooling rate (Eq.~(\ref{eq:P_t})), even when some suppression due to ionization of the nugget itself is taken into account, ensures that the majority of this energy is deposited within the first $0.1 - 0.2~\rm s$. This implies that the dominant portion of the energy from Eg.~(\ref{AQN-accumulated}) will be deposited at altitudes below $z_0 \approx 50~\rm km$ assuming the typical DM velocity of $\sim 250~\rm km/s$. While AQNs continue to emit radiation after leaving the atmosphere, the intensity diminishes significantly compared to their initial upward movement.}:
\beq
 \label{AQN-accumulated}
   E_{\rm AQN}(T) &=&\int c_VdT\approx       \frac{T^2(\mu_d^2+\mu_u^2)}{6}   \frac{4\pi R^3}{3}  \\
   &\approx& 10^{10}J \cdot \left(\frac{T}{400 ~ \rm keV}\right)^2\cdot \left(\frac{\mu_{u,d}}{500 ~ \rm MeV}\right)^2. \nonumber
\eeq

To estimate the total energy deposited by all AQNs hitting the Earth, we multiply the event rate from Eq.~(\ref{Phi1}) by the energy deposited per single event from Eq.~(\ref{AQN-accumulated}):
\be
 \label{energy-deposition}
  \frac{dE}{dt dA}  \approx     \frac{0. 1~W}{m^2} \cdot \left(\frac{A(t)}{10^4}\right)\cdot    \left(\frac{T}{400 ~{\rm keV}}\right)^{2},
 \ee
where $A(t)$ is the enhancement factor defined in Eq.~(\ref{enhancement}), accounting for the ``streaming invisible matter" hypothesis. This factor plays a central role in the proposal, as it reflects deviations from the SHM for local DM behavior, as discussed in \cite{Bertolucci:2016xjm, Zioutas:2020ndf, Zioutas:2023ybw, Maroudas:2022ufl}.

Taking Eq.~(\ref{energy-deposition}) at face value while assuming the parameter $A(t)$ having the value from Eq.~(\ref{enhancement}) we find that the estimated energy deposition rate, $\sim 0.1~\rm W/m^2$, aligns well with the observational constraint in Eq.~(\ref{E-deposition}) which in turn was derived from the mysterious observations (items \ref{item:first}-\ref{item:fifth}). Therefore, this numerical consistency strengthens the plausibility of our conjecture.

Encouraged by this  numerical  result, we propose the following identification:
  \be
  \label{eq:proposal}
  \boldmath
 \rm ``streaming~ invisible ~matter" ~\equiv ~[AQNs],
 \ee
This equivalence suggests that the mysterious ``streaming invisible matter" conjectured in \cite{Bertolucci:2016xjm, Zioutas:2020ndf, Zioutas:2023ybw, Maroudas:2022ufl} corresponds to AQNs. In other words, we offer a microscopic mechanism capable of explaining the puzzling observations (items \ref{item:first}-\ref{item:fifth}).

In Sect.~\ref{sect:consistency}, we present several arguments demonstrating that the proposal in Eq.~(\ref{eq:proposal}) is indeed consistent with the observed anomalies as reported in \cite{Bertolucci:2016xjm, Zioutas:2020ndf, Zioutas:2023ybw, Maroudas:2022ufl}.

\section{Proposal (\ref{eq:proposal}) confronts the observations}
\label{sect:consistency}

The main objective of this section is to demonstrate that the mysterious puzzles introduced in Sect.~\ref{mysteries} are naturally and qualitatively resolved within the framework of the proposal (\ref{eq:proposal}). We address each puzzle in the same order as presented earlier.

\subsection{Puzzle 1: Stratospheric temperature variations}

The striking temperature excursions observed annually in December-January at altitudes of $38-47~\rm km$ (see Fig.~1 in \cite{Zioutas:2020ndf}) present a significant anomaly superimposed on the otherwise smooth seasonal variations (see item \ref{item:first}). Additionally, these variations, which can not be produced by any known phenomenon intrinsic to the atmosphere, become less pronounced at lower and higher atmospheric regions.

Within the AQN framework proposed in (\ref{eq:proposal}), this pattern emerges naturally. Upward-moving AQNs lose their accumulated heat as they propagate through the atmosphere. Consequently, their internal temperature and the rate of energy deposition decrease significantly at higher altitudes, as explained in Sect.~\ref{AQN-dense} and footnote \ref{altitude}.

At lower altitudes, while the energy deposition is much higher due to the increased air density $n (z) \approx (- {z}/{h}) $, the temperature change $\Delta T (z\approx 0)$ is less pronounced because the denser air distributes the deposited energy more efficiently. Assuming the energy injection is uniform along the AQN's path up to a cutoff altitude $z\lesssim 50~\rm km$, and that the spectral features of the radiation are approximately the same, while the thermalization processes are consistent across altitudes, this relationship can be qualitatively expressed as:
\be
 \label{eq:altitude}
 \Delta T (z\approx 0)\propto  {\Delta T (z)}\exp{\left(-\frac{z}{h}\right)}, ~~~ h~\approx 8~ \rm km, 
 \ee
where $\Delta T(z \approx 0)$ represents the temperature change at sea level. This equation indicates that, for the same energy deposition, temperature variations $\Delta T$ are less pronounced at lower altitudes due to the denser atmosphere.

While the above mechanism explains the altitude dependence qualitatively, there are additional effects that influence the AQN's behavior. For instance, AQNs temperature will increase due to annihilation processes in the atmosphere. This additional heating is stronger at lower altitudes where the density $n$ is higher (see Eq.~(\ref{T})). However, this contribution is expected to be much smaller than the energy in Eq.~(\ref{AQN-accumulated}) accumulated by AQNs during their passage through the Earth's interior. Thus, while the expected sharp dependence on altitude $z\lesssim z_0$ described by Eq.~(\ref{eq:altitude}) will be considerably reduced, the overall trend should remain valid qualitatively. The reduced temperature effect at lower altitudes and the diminished energy deposition at higher altitudes together naturally reproduce the observed patterns.

\subsection{Puzzle 2: Solar non-correlation}

The second puzzle (see item \ref{item:second}) concerns the lack of correlation between the observed $\Delta T$ variations and solar activity. Given the Sun's dominant role in driving Earth's atmospheric dynamics, it is natural to suspect it as the source of the observed temperature anomalies. However, as \cite{Zioutas:2020ndf, Maroudas:2022ufl} shows, this correlation is absent. Within the AQN framework, this non-correlation is naturally resolved. AQNs, as DM particles, are independent of solar activity and their interactions with the Earth's atmosphere are unaffected by variations in solar irradiation. Consequently, the lack of a connection between $\Delta T$ variations and solar activity follows directly from the AQN hypothesis.

\subsection{Puzzle 3: Planetary correlation}

The third puzzle (see item \ref{item:third}) is a central element of the analysis in \cite{Zioutas:2020ndf}, which identified anomalous correlations between planetary positions and temperature variations. Within our framework, the conjecture from Eq.~(\ref{eq:proposal}) provides a natural explanation by identifying the ``streaming invisible matter” proposed in \cite{Zioutas:2020ndf} with slow-moving AQN DM particles. The AQNs satisfy the condition from Eq.~(\ref{sigma/m}) for DM candidates and are capable of generating substantial energy deposition, as shown in Eq.~(\ref{energy-deposition}). Moreover, due to their slow speed, streaming AQNs are subject to gravitational lensing, making them ideal candidates for the hypothesized ``streaming invisible matter" responsible for the planetary correlations observed in \cite{Zioutas:2020ndf, Maroudas:2022ufl}.

\subsection{Puzzle 4: TEC}

The fourth puzzle (see item \ref{item:fourth}) involves anomalies in the ionization of the Earth's entire atmosphere (i.e. mainly its ionosphere) which plays the role of a huge detector measuring the TEC. The observed correlations between TEC and planetary positions \cite{Bertolucci:2016xjm}are naturally explained within the AQN framework if accepting Eq.~(\ref{eq:proposal}).

As AQNs propagate through the stratosphere, they cool down while they emit energetic photons, predominantly in the EUV and X-ray frequency bands, due to bremsstrahlung radiation from the electrosphere being very flat for $\omega\lesssim T$\cite{Forbes:2008uf}. These photons play two key roles.
First, they contribute to local temperature increases along the AQN path as a result of thermalization, addressing Puzzle 1. Secondly, the same AQN-induced EUV and X-ray photons will ionize the neutral atoms and molecules when AQN propagates in the atmosphere, therefore contributing to TEC. Thus, the observed TEC variations arise from the same AQN-induced flux that drives the temperature anomalies, with both phenomena sharing a common origin in the energy deposition described by Eq.~(\ref{energy-deposition}).

Furthermore, the AQN-induced spectrum is dominated by EUV and X-ray radiation with the intensity from Eq.~(\ref{energy-deposition}) being more than sufficient to explain TEC anomalies. Variations of similar or even lower intensity in these frequency bands are known to drive TEC changes during the conventional 11-year solar cycles. Therefore, within the AQN framework, $\Delta T$ and TEC must exhibit correlated planetary dependence, consistent with the observations in \cite{Zioutas:2020ndf}.

\subsection{Puzzle 5: TEC-Earthquake correlation}

The fifth puzzle (see item \ref{item:fifth}) is particularly enigmatic as it involves observed correlations between TEC variations in the ionosphere and large earthquakes originating deep within the Earth's interior. Establishing a connection between these phenomena is challenging, as electromagnetic (EM) radiation, which drives TEC, has minimal coupling to acoustic disturbances capable of triggering earthquakes. The energy deposition from AQNs (Eq.~(\ref{energy-deposition})), though significant, is four orders of magnitude smaller than the solar constant ($\sim 1400\rm W/m^2$) representing the power we continuously receive on Earth from Sun, which does not generate similar correlations with earthquakes. However, insights can be gained by examining analogous correlations between solar flares and sunquakes.

\subsubsection{Lessons from the observed correlation of solar flares and Sunquakes}
\label{Sun}

It has been established that sunquake events are well correlated with hard X-ray emissions during the impulsive phase of solar flares. Conventional explanations suggest that flare energy, released in the corona, drives acoustic disturbances in the solar interior near the photosphere. However, this mechanism requires an energy propagation across nine pressure scale heights, a process that is difficult to reconcile with standard astrophysical models \cite{Judge_2014,2015ApJ...812...35M}.

Authors in \cite{Judge_2014} hypothesize that ``the energy is transported downwards in a fashion that is somehow {\it invisible} to our observations." This observation aligns naturally with the AQN framework proposed in \cite{Zhitnitsky:2018mav}, where AQNs are hypothesized to trigger both solar flares and sunquakes. Specifically, they can first initiate the large solar flares in the solar corona, and secondly, since they can easily propagate downwards and penetrate to very deep regions of the solar photosphere, they can play the role of the triggers initiating the sunquakes in the photosphere 

In the AQN model, nuggets entering the solar atmosphere possess high velocities ($v_{\rm AQN} \sim 700~\rm km/s$) and large Mach numbers ($M\equiv \frac{v_{AQN}}{c_s} \gtrsim 10 $). This generates strong shock waves, with discontinuities in temperature and pressure scaling as \cite{Zhitnitsky:2018mav}:
\beq
 \label{shock1}
    \frac{p_2}{p_1}\simeq M^2\cdot \frac{2\gamma }{(\gamma+1)}, ~~~~ \frac{T_2}{T_1}\simeq M^2\cdot \frac{2\gamma(\gamma-1) }{(\gamma+1)^2},
 \eeq
where $\gamma = 5/3$ is the specific heat ratio and the leading $M^2\gg 1$ terms are kept. These relations imply that strong shock waves, driven by AQNs with $M \gg 1$, can trigger localized disturbances deep within the dense region of the photosphere, igniting sunquakes. This is because the shock wave generated due to the large Mach number may produce a single highly localized increase of the temperature $\Delta T/T\gg 1$ and pressure $\Delta p/p\gg1 $ deep in the atmosphere where the sunquakes originate. 

This mechanism also explains why sunquakes are localized within flaring regions: the disturbance is directly tied to the AQN's entry point in the photosphere, not the overall flare energy. Furthermore, large Mach numbers $M\gg 1$ do not guarantee that a large flare will be developed. The AQN must enter the active region with strong magnetic fields, in which case a large flare indeed can then be ignited by the AQNs \cite{Zhitnitsky:2018mav}. In regions of quiet Sun, AQNs deposit energy as EUV and X-ray radiation without triggering large flares, as the area being hit must satisfy certain conditions to ignite a large solar flare\footnote{Necessary, but not sufficient condition is the presence of a large magnetic field in the active region such that the magnetic reconnection may become operational.}.

\subsubsection{On the puzzling correlations between TEC and Earthquakes}
\label{Earth}

We now can address the observed correlation between TEC variations in the ionosphere and large earthquakes from a microscopical perspective. The key insight lies in the Mach number of AQNs moving beneath the Earth's surface, which we estimate as:
 \be
 \label{Mach}
 M\equiv \frac{v_{AQN}}{c_s} \approx \frac{250~ \rm km/s}{8~\rm km/s}\approx 30\gg 1,
  \ee    
where we assume the speed of sound in solid rock is approximately $8~\rm km/s$. This large Mach number implies that, similar to the Sun's case mentioned before, shock waves will inevitably form as AQNs propagate through the Earth's interior (whether through water or solid rock). These shock waves generate discontinuities in temperature $\Delta T/T$ and pressure $\Delta p/p$ along the AQN's path which will be numerically enormously large.

For example, the overpressure generated by the blast wave in solid rock has been estimated at $\Delta p \sim 10^7~\rm Pa$ \cite{Budker:2020mqk}. If an AQN traverses a seismically active region, it can serve as a trigger for a large earthquake. It is well known that earthquakes can be induced by human activities such as nuclear tests. Similarly, an AQN—by generating significant overpressure along its trajectory—can initiate an earthquake in regions where only a ``small push” is required to trigger a major event.

This mechanism closely parallels the role of AQNs in initiating solar flares and sunquakes (discussed in Sect.~\ref{Sun}). In both cases, the AQN acts as a trigger, initiating a much larger event powered by the internal physics of the medium (the Earth's crust or the Sun), rather than the AQN itself.

Therefore, the correlation between TEC variations and earthquake rates, as reported in \cite{Zioutas:2023ybw, Zioutas_geoscience}, arises naturally within the AQN framework because both phenomena are linked to the same AQNs traversing the Earth and its atmosphere. Both phenomena are thus proportional to the AQN event rate (Eq.~(\ref{Phi1})) and the energy deposition rate (Eq.~(\ref{energy-deposition})), with the enhancement factor $A(t)$ (Eq.~(\ref{enhancement})) playing a critical role. As $A(t)$ increases spatiotemporally due to gravitational focusing by the solar system bodies, reflecting the flux of ``streaming invisible matter," the rates of both TEC variations and earthquakes are expected to rise.

It is important to emphasize that, in this framework, AQNs act as triggers for earthquakes rather than TEC variations directly influencing seismic activity. This mirrors the observed correlations between sunquakes and flares in the Sun (Sect.~\ref{Sun}), where both phenomena are initiated by the same AQN. In the case of the Sun, such correlations are easier to observe because flares and sunquakes are short-lived (lasting only hours) and well-localized in space, making their relationship straightforward to establish. On Earth, the correlations between TEC and earthquakes are more challenging to study due to the longer timescales and broader spatial extent of seismic events.

In conclusion, the puzzles introduced in Sect.~\ref{mysteries} appear far less mysterious when viewed through the lens of the AQN framework. Instead, they emerge as natural consequences of the same underlying mechanism. In the next section, we discuss possible tests of this framework and propose experiments to evaluate the specific conjecture (\ref{eq:proposal}).

\section{Conclusion and Outlook} 
\label{conclusion}

The presence of antimatter nuggets in the AQN framework implies an abundance of observable effects across a wide range of scales, from the early Universe to galactic structures, the Sun, and terrestrial phenomena. As discussed in Sect.~\ref{AQN}, the antimatter nuggets naturally arise as a resolution to the long-standing puzzle of the near-equal densities of visible matter and DM, $\Omega_{\rm DM} \sim \Omega_{\rm visible}$. This remarkable feature is rooted in the dynamics of the $\cal{CP}$-violating axion field during the QCD formation period.

In this work, we focused on the implications of AQN annihilation events for resolving the mysterious puzzles \ref{item:first}-\ref{item:fifth} by proposing specific mechanisms to explain these phenomena, supported by observational data. In the following subsections, we discuss potential tests for our proposal and outline broader implications of the AQN framework for puzzles across different scales.

\subsection{Possible tests of the proposal}
\label{sect:tests}

The AQN framework offers a promising microscopical mechanism to explain the observed correlations and anomalies discussed in Sect.~\ref{mysteries}. A key feature of these studies is that large-scale regions of the Earth, including the stratosphere, ionosphere, and seismically active zones, effectively act as ``natural detectors" for AQN-induced events. However, dedicated instruments are essential for direct detection and validation of this fundamentally new type of strongly interacting DM.

A primary challenge in studying AQN-induced phenomena is distinguishing DM signals from significant background noise and unrelated events. A promising strategy to overcome this, as suggested in \cite{Budker:2019zka, Liang:2020mnz}, is to use networks of synchronized detectors. By correlating time delays between two or more detectors, one can isolate AQN-induced signals based on their unique propagation characteristics, since these time delays are unambiguously fixed by the distances between the instruments.

When AQNs propagate and annihilate in the Earth's atmosphere or interior, they emit relativistic axions \footnote{The emitted axions in the AQN framework have velocities $v_a \approx 0.6c$ which are detectable via broadband instruments such as WISPLC \cite{Zhang:2021bpa}, ABRACADABRA \cite{Kahn:2016aff}, LC Circuit \cite{Sikivie:2013laa}, DM Radio \cite{Chaudhuri:2018rqn}, MADMAX \cite{Gardikiotis:2024crt} and BRASS \cite{Bajjali:2023uis}. See also the recent reviews \cite{Marsh:2015xka, Graham:2015ouw, Battesti:2018bgc, Irastorza:2018dyq, Sikivie:2020zpn} with mentioning existing and planning instruments which are capable of working in a broadband mode. See also } and neutrinos with very large mean free paths. They also generate infrasound waves that can propagate to very large distances which can be detectable by Distributed Acoustic Sensing (DAS) systems \cite{Budker:2020mqk}. These signals exhibit short-lived spikes, particularly in areas where AQNs enter or exit the Earth’s surface. The typical rate for such correlated events depends on the strength of the spikes as given in Table IV in \cite{Liang:2019lya}.

As an example, the event rate derived in Eq.~(\ref{Phi1}) suggests that AQNs impact the Earth’s surface approximately once per day within a $(100\mathrm{km}^2)$ area. Such events are expected to produce signal spikes with an enhancement factor of $A(t) \sim 10^2$. If two or more detectors are situated within this area and separated by distances $d \lesssim 100~\mathrm{km}$, one can anticipate well-defined time delays $\tau$ between detections, determined by the AQN velocity $v_{\rm AQN}$, the speed of sound $c_s$, and the geometry of the network. Furthermore, the synchronization between different instruments could play a vital role in the discovery of the DM in the form of AQNs.

Additionally, a network of detectors can allow a study of the directionality of the DM in the form of the AQNs \cite{Liang:2020mnz} by testing the local features of the DM distribution which are expected to be drastically different from SHM as expressed by Eq.~(\ref{enhancement}). Therefore, this will ultimately enable a test of the ``streaming invisible matter'' hypothesis from \cite{Bertolucci:2016xjm, Zioutas:2020ndf, Zioutas:2023ybw, Maroudas:2022ufl}.

\subsection{Other (indirect) evidence for DM in the form of AQNs}
\label{sect:paradigm}

There are numerous hints suggesting that annihilation events—an inevitable feature of this framework—may have occurred during the early Universe, the epoch of galaxy formation, and even in the present day. These events, induced by AQN interactions with visible matter, could provide explanations for a range of unresolved puzzles across different scales.

We begin with a longstanding mystery from the early Universe: the ``Primordial Lithium Puzzle.'' This problem has persisted for decades, but it has been argued in \cite{Flambaum:2018ohm} that AQNs during the BBN epoch do not affect the production of H and He. However, they might resolve the lithium anomaly due to the large electric charge ($Z=3$) of Li, which interacts strongly with the negatively\footnote{The AQN will be ionized as a result of high temperature during this epoch when many weakly bound positrons from the AQN's electrosphere leave the system such that the AQN becomes strongly negatively charged object during BBN epoch.} charged AQNs.

Other well-known puzzles are related to galaxy formation. The most commonly expressed ones are the ``Core-Cusp Problem", the ``Missing Satellite Problem" and the ``Too-Big-to-Fail Problem". These and other challenges to the conventional understanding of galactic structure are detailed in \cite{Tulin:2017ara, Salucci:2020eqo}. It has been proposed in \cite{Zhitnitsky:2023znn} that these discrepancies may be alleviated if DM exists in the form of composite, nuclear-density objects within the AQN framework.

Shifting to present-day observations, another significant puzzle concerns the diffuse UV emission in our galaxy. As discussed in \cite{Henry_2014, Akshaya_2018, 2019MNRAS.489.1120A}, several observations challenge the conventional explanation that the dominant source of diffuse UV background is dust-scattered radiation from UV-emitting stars. First, the diffuse UV radiation is remarkably uniform across both hemispheres, in stark contrast to the uneven distribution of UV-emitting stars. Second, it shows almost no dependence on the Galactic longitude, which is inconsistent with the observed confinement of bright UV stars to longitudes between $180^\circ$ and $360^\circ$. These anomalies strongly suggest that the diffuse UV radiation cannot originate from starlight. The authors of \cite{Henry_2014} even describe the source of this radiation as ``unknown". It has been proposed in \cite{Zhitnitsky:2021wjb} that this UV radiation excess could be explained by AQN annihilation events. This proposal is supported by demonstrating that the intensity and spectral features of AQN-induced emissions align closely with the observed characteristics of the UV excess reported in \cite{Henry_2014, Akshaya_2018, 2019MNRAS.489.1120A}.

At the solar scale, AQNs may also offer an explanation for the renowned ``Solar Corona Mystery", a long-standing problem\footnote{the quiet Sun with a magnetic field $B \sim 1$ Gauss emits EUV radiation at photon energies of $\sim 100~\rm eV$, a phenomenon unexplained by conventional astrophysics. This occurs within an atmospheric layer only $\sim 100$ km thick, where the temperature steeply rises to several million kelvins. The variation of EUV with the solar cycle is modest (20-30\%), while the magnetic activity varies by a factor of 100 or more. Therefore it seems implausible that magnetic reconnection alone, which is known to be responsible for large flares, could account for this emission when $B \sim 1$ Gauss. There are other puzzling solar features, discussed in \cite{Zhitnitsky:2017rop, Raza:2018gpb}.} involving the anomalous behavior of the Sun. Specifically, the ``nanoflares" hypothesized by Parker \cite{Parker} could be identified as annihilation events caused by AQNs \cite{Zhitnitsky:2017rop, Raza:2018gpb}.

Finally, we turn our attention to Earth. There are growing indications that our current understanding of ultra-high-energy cosmic rays (UHECR) — including their sources, nature, and propagation — remains incomplete or possibly flawed. This perspective is supported by the CREDO (Cosmic-Ray Extremely Distributed Observatory) collaboration, which has compiled a compelling list of arguments highlighting these gaps \cite{sym12111835}. It is worth considering that some events typically interpreted as UHECR phenomena might instead be misidentified or misclassified, as suggested in Sect.~\ref{aqn_hypothesis}. These puzzling events could be manifestations of AQN-induced effects rather than conventional cosmic-ray behavior.

A key proposal arising from this consideration involves combining detection methods: placing an acoustic instrument, as mentioned in Sect.~\ref{aqn_hypothesis}, alongside a CR detector in the same geographical area. This approach would enable to distinguish genuine UHECR events from AQN-induced phenomena. Such a capability would not only refine the analysis of UHECR but also align with the objectives of the CREDO collaboration, which seeks to challenge and enhance conventional models of cosmic-ray physics. Moreover, this methodology could enable a search for correlations between earthquakes and CR-like AQN-induced events. As argued in Sect.~\ref{sect:consistency}, such a correlation would naturally arise if both phenomena are driven by the same flux of ``streaming invisible matter" proposed in the AQN framework. This dual-detection strategy would offer a powerful tool for exploring and validating the broader implications of the AQN hypothesis.

In conclusion, this work advocates a paradigm shift in the understanding of DM: from weakly interacting non-baryonic particles to strongly interacting baryonic composite objects made from (anti)quarks and gluons of the Standard Model (see Sect.~\ref{AQN}). The AQN model is consistent with all current observations across cosmological, astrophysical, and terrestrial scales. Moreover, it offers potential resolutions to long-standing mysteries, from the early Universe to the present day. If validated through the proposed tests and experiments, this framework could revolutionize our understanding of DM and its role in the cosmos.

\section*{Acknowledgements}
The motivation for this work emerged as a result of long, emotional, and never-ending discussions with  Konstantin Zioutas. A.Z is very thankful to him for these discussions. 
Part of this project is funded by the Deutsche Forschungsgemeinschaft (DFG, German Research Foundation) under Germany's Excellence Strategy -  EXC 2121 ``Quantum Universe"- 390833306, and through the DFG funds for major instrumentation grant DFG INST 152/8241. The research  for AZ was supported in part by the Natural Sciences and Engineering Research Council of Canada.

\appendix

\bibliography{references}

\begin{thebibliography}{75}%
\makeatletter
\providecommand \@ifxundefined [1]{%
 \@ifx{#1\undefined}
}%
\providecommand \@ifnum [1]{%
 \ifnum #1\expandafter \@firstoftwo
 \else \expandafter \@secondoftwo
 \fi
}%
\providecommand \@ifx [1]{%
 \ifx #1\expandafter \@firstoftwo
 \else \expandafter \@secondoftwo
 \fi
}%
\providecommand \natexlab [1]{#1}%
\providecommand \enquote  [1]{``#1''}%
\providecommand \bibnamefont  [1]{#1}%
\providecommand \bibfnamefont [1]{#1}%
\providecommand \citenamefont [1]{#1}%
\providecommand \href@noop [0]{\@secondoftwo}%
\providecommand \href [0]{\begingroup \@sanitize@url \@href}%
\providecommand \@href[1]{\@@startlink{#1}\@@href}%
\providecommand \@@href[1]{\endgroup#1\@@endlink}%
\providecommand \@sanitize@url [0]{\catcode `\\12\catcode `\$12\catcode
  `\&12\catcode `\#12\catcode `\^12\catcode `\_12\catcode `\%12\relax}%
\providecommand \@@startlink[1]{}%
\providecommand \@@endlink[0]{}%
\providecommand \url  [0]{\begingroup\@sanitize@url \@url }%
\providecommand \@url [1]{\endgroup\@href {#1}{\urlprefix }}%
\providecommand \urlprefix  [0]{URL }%
\providecommand \Eprint [0]{\href }%
\providecommand \doibase [0]{http://dx.doi.org/}%
\providecommand \selectlanguage [0]{\@gobble}%
\providecommand \bibinfo  [0]{\@secondoftwo}%
\providecommand \bibfield  [0]{\@secondoftwo}%
\providecommand \translation [1]{[#1]}%
\providecommand \BibitemOpen [0]{}%
\providecommand \bibitemStop [0]{}%
\providecommand \bibitemNoStop [0]{.\EOS\space}%
\providecommand \EOS [0]{\spacefactor3000\relax}%
\providecommand \BibitemShut  [1]{\csname bibitem#1\endcsname}%
\let\auto@bib@innerbib\@empty
\bibitem [{\citenamefont {Bertolucci}\ \emph {et~al.}(2017)\citenamefont
  {Bertolucci}, \citenamefont {Zioutas}, \citenamefont {Hofmann},\ and\
  \citenamefont {Maroudas}}]{Bertolucci:2016xjm}%
  \BibitemOpen
  \bibfield  {author} {\bibinfo {author} {\bibfnamefont {S.}~\bibnamefont
  {Bertolucci}}, \bibinfo {author} {\bibfnamefont {K.}~\bibnamefont {Zioutas}},
  \bibinfo {author} {\bibfnamefont {S.}~\bibnamefont {Hofmann}}, \ and\
  \bibinfo {author} {\bibfnamefont {M.}~\bibnamefont {Maroudas}},\ }\href
  {\doibase 10.1016/j.dark.2017.06.001} {\bibfield  {journal} {\bibinfo
  {journal} {Phys. Dark Univ.}\ }\textbf {\bibinfo {volume} {17}},\ \bibinfo
  {pages} {13} (\bibinfo {year} {2017})},\ \Eprint
  {http://arxiv.org/abs/1602.03666} {arXiv:1602.03666 [astro-ph.SR]}
  \BibitemShut {NoStop}%
\bibitem [{\citenamefont {Zioutas}\ \emph {et~al.}(2020)\citenamefont
  {Zioutas}, \citenamefont {Argiriou}, \citenamefont {Fischer}, \citenamefont
  {Hofmann}, \citenamefont {Maroudas}, \citenamefont {Pappa},\ and\
  \citenamefont {Semertzidis}}]{Zioutas:2020ndf}%
  \BibitemOpen
  \bibfield  {author} {\bibinfo {author} {\bibfnamefont {K.}~\bibnamefont
  {Zioutas}}, \bibinfo {author} {\bibfnamefont {A.}~\bibnamefont {Argiriou}},
  \bibinfo {author} {\bibfnamefont {H.}~\bibnamefont {Fischer}}, \bibinfo
  {author} {\bibfnamefont {S.}~\bibnamefont {Hofmann}}, \bibinfo {author}
  {\bibfnamefont {M.}~\bibnamefont {Maroudas}}, \bibinfo {author}
  {\bibfnamefont {A.}~\bibnamefont {Pappa}}, \ and\ \bibinfo {author}
  {\bibfnamefont {Y.~K.}\ \bibnamefont {Semertzidis}},\ }\href {\doibase
  10.1016/j.dark.2020.100497} {\bibfield  {journal} {\bibinfo  {journal} {Phys.
  Dark Univ.}\ }\textbf {\bibinfo {volume} {28}},\ \bibinfo {pages} {100497}
  (\bibinfo {year} {2020})},\ \Eprint {http://arxiv.org/abs/2004.11006}
  {arXiv:2004.11006 [astro-ph.EP]} \BibitemShut {NoStop}%
\bibitem [{\citenamefont {Zioutas}\ \emph {et~al.}(2023)\citenamefont {Zioutas}
  \emph {et~al.}}]{Zioutas:2023ybw}%
  \BibitemOpen
  \bibfield  {author} {\bibinfo {author} {\bibfnamefont {K.}~\bibnamefont
  {Zioutas}} \emph {et~al.}\ }(\bibinfo {year} {2023})\ \Eprint
  {http://arxiv.org/abs/2309.10779} {arXiv:2309.10779 [astro-ph.EP]}
  \BibitemShut {NoStop}%
\bibitem [{\citenamefont {{Zhitnitsky}}(2003)}]{Zhitnitsky:2002qa}%
  \BibitemOpen
  \bibfield  {author} {\bibinfo {author} {\bibfnamefont {A.~R.}\ \bibnamefont
  {{Zhitnitsky}}},\ }\href {\doibase 10.1088/1475-7516/2003/10/010} {\bibfield
  {journal} {\bibinfo  {journal} {\jcap}\ }\textbf {\bibinfo {volume} {10}},\
  \bibinfo {eid} {010} (\bibinfo {year} {2003})},\ \Eprint
  {http://arxiv.org/abs/hep-ph/0202161} {hep-ph/0202161} \BibitemShut {NoStop}%
\bibitem [{\citenamefont
  {Zhitnitsky}(2021{\natexlab{a}})}]{Zhitnitsky:2021iwg}%
  \BibitemOpen
  \bibfield  {author} {\bibinfo {author} {\bibfnamefont {A.}~\bibnamefont
  {Zhitnitsky}},\ }\href {\doibase 10.1142/S0217732321300172} {\bibfield
  {journal} {\bibinfo  {journal} {Mod. Phys. Lett. A}\ }\textbf {\bibinfo
  {volume} {36}},\ \bibinfo {pages} {2130017} (\bibinfo {year}
  {2021}{\natexlab{a}})},\ \Eprint {http://arxiv.org/abs/2105.08719}
  {arXiv:2105.08719 [hep-ph]} \BibitemShut {NoStop}%
\bibitem [{\citenamefont {Maroudas}(2022)}]{Maroudas:2022ufl}%
  \BibitemOpen
  \bibfield  {author} {\bibinfo {author} {\bibfnamefont {M.}~\bibnamefont
  {Maroudas}},\ }\emph {\bibinfo {title} {{Signals for invisible matter from
  solar - terrestrial observations}}},\ \href {\doibase 10.12681/eadd/51922}
  {Ph.D. thesis} (\bibinfo {year} {2022}),\ \Eprint
  {http://arxiv.org/abs/2404.02290} {arXiv:2404.02290 [hep-ex]} \BibitemShut
  {NoStop}%
\bibitem [{\citenamefont {Zioutas}\ \emph {et~al.}(2022)\citenamefont
  {Zioutas}, \citenamefont {Maroudas},\ and\ \citenamefont
  {Kosovichev}}]{Zioutas_2022_Radius}%
  \BibitemOpen
  \bibfield  {author} {\bibinfo {author} {\bibfnamefont {K.}~\bibnamefont
  {Zioutas}}, \bibinfo {author} {\bibfnamefont {M.}~\bibnamefont {Maroudas}}, \
  and\ \bibinfo {author} {\bibfnamefont {A.}~\bibnamefont {Kosovichev}},\
  }\href {\doibase 10.3390/sym14020325} {\bibfield  {journal} {\bibinfo
  {journal} {Symmetry}\ }\textbf {\bibinfo {volume} {14}} (\bibinfo {year}
  {2022}),\ 10.3390/sym14020325}\BibitemShut {NoStop}%
\bibitem [{\citenamefont {Hoffmann}\ \emph {et~al.}(2003)\citenamefont
  {Hoffmann}, \citenamefont {Jacoby},\ and\ \citenamefont
  {Zioutas}}]{Hoffmann_2003_gravitational}%
  \BibitemOpen
  \bibfield  {author} {\bibinfo {author} {\bibfnamefont {D.}~\bibnamefont
  {Hoffmann}}, \bibinfo {author} {\bibfnamefont {J.}~\bibnamefont {Jacoby}}, \
  and\ \bibinfo {author} {\bibfnamefont {K.}~\bibnamefont {Zioutas}},\ }\href
  {\doibase https://doi.org/10.1016/S0927-6505(03)00138-5} {\bibfield
  {journal} {\bibinfo  {journal} {Astroparticle Physics}\ }\textbf {\bibinfo
  {volume} {20}},\ \bibinfo {pages} {73} (\bibinfo {year} {2003})}\BibitemShut
  {NoStop}%
\bibitem [{\citenamefont {Kryemadhi}\ \emph {et~al.}(2023)\citenamefont
  {Kryemadhi}, \citenamefont {Maroudas}, \citenamefont {Mastronikolis},\ and\
  \citenamefont {Zioutas}}]{Kryemadhi_2023_gravitational}%
  \BibitemOpen
  \bibfield  {author} {\bibinfo {author} {\bibfnamefont {A.}~\bibnamefont
  {Kryemadhi}}, \bibinfo {author} {\bibfnamefont {M.}~\bibnamefont {Maroudas}},
  \bibinfo {author} {\bibfnamefont {A.}~\bibnamefont {Mastronikolis}}, \ and\
  \bibinfo {author} {\bibfnamefont {K.}~\bibnamefont {Zioutas}},\ }\href
  {\doibase 10.1103/PhysRevD.108.123043} {\bibfield  {journal} {\bibinfo
  {journal} {Phys. Rev. D}\ }\textbf {\bibinfo {volume} {108}},\ \bibinfo
  {pages} {123043} (\bibinfo {year} {2023})}\BibitemShut {NoStop}%
\bibitem [{\citenamefont {Cantatore}\ \emph {et~al.}(2023)\citenamefont
  {Cantatore}, \citenamefont {\c{C}etin}, \citenamefont {Fischer},
  \citenamefont {Funk}, \citenamefont {Karuza}, \citenamefont {Kryemadhi},
  \citenamefont {Maroudas}, \citenamefont {\"Ozbozduman}, \citenamefont
  {Semertzidis},\ and\ \citenamefont {Zioutas}}]{Cantatore:2020obc}%
  \BibitemOpen
  \bibfield  {author} {\bibinfo {author} {\bibfnamefont {G.}~\bibnamefont
  {Cantatore}}, \bibinfo {author} {\bibfnamefont {S.~A.}\ \bibnamefont
  {\c{C}etin}}, \bibinfo {author} {\bibfnamefont {H.}~\bibnamefont {Fischer}},
  \bibinfo {author} {\bibfnamefont {W.}~\bibnamefont {Funk}}, \bibinfo {author}
  {\bibfnamefont {M.}~\bibnamefont {Karuza}}, \bibinfo {author} {\bibfnamefont
  {A.}~\bibnamefont {Kryemadhi}}, \bibinfo {author} {\bibfnamefont
  {M.}~\bibnamefont {Maroudas}}, \bibinfo {author} {\bibfnamefont
  {K.}~\bibnamefont {\"Ozbozduman}}, \bibinfo {author} {\bibfnamefont {Y.~K.}\
  \bibnamefont {Semertzidis}}, \ and\ \bibinfo {author} {\bibfnamefont
  {K.}~\bibnamefont {Zioutas}},\ }\href {\doibase 10.3390/sym15061167}
  {\bibfield  {journal} {\bibinfo  {journal} {Symmetry}\ }\textbf {\bibinfo
  {volume} {15}},\ \bibinfo {pages} {1167} (\bibinfo {year} {2023})},\ \Eprint
  {http://arxiv.org/abs/2012.03353} {arXiv:2012.03353 [hep-ex]} \BibitemShut
  {NoStop}%
\bibitem [{\citenamefont {Zioutas}\ \emph {et~al.}(2021)\citenamefont {Zioutas}
  \emph {et~al.}}]{Zioutas:2021xcm}%
  \BibitemOpen
  \bibfield  {author} {\bibinfo {author} {\bibfnamefont {K.}~\bibnamefont
  {Zioutas}} \emph {et~al.},\ }\href {\doibase 10.3390/ECU2021-09313}
  {\bibfield  {journal} {\bibinfo  {journal} {Phys. Sc. Forum}\ }\textbf
  {\bibinfo {volume} {2}},\ \bibinfo {pages} {10} (\bibinfo {year} {2021})},\
  \Eprint {http://arxiv.org/abs/2108.11647} {arXiv:2108.11647 [astro-ph.SR]}
  \BibitemShut {NoStop}%
\bibitem [{\citenamefont {Lazanu}\ and\ \citenamefont
  {Parvu}(2024)}]{Lazanu:2024ddm}%
  \BibitemOpen
  \bibfield  {author} {\bibinfo {author} {\bibfnamefont {I.}~\bibnamefont
  {Lazanu}}\ and\ \bibinfo {author} {\bibfnamefont {M.}~\bibnamefont {Parvu}},\
  }\href {\doibase 10.1088/1475-7516/2024/05/014} {\bibfield  {journal}
  {\bibinfo  {journal} {JCAP}\ }\textbf {\bibinfo {volume} {05}},\ \bibinfo
  {pages} {014} (\bibinfo {year} {2024})},\ \Eprint
  {http://arxiv.org/abs/2402.07312} {arXiv:2402.07312 [hep-ph]} \BibitemShut
  {NoStop}%
\bibitem [{\citenamefont {Zioutas}\ \emph
  {et~al.}(2024{\natexlab{a}})\citenamefont {Zioutas} \emph
  {et~al.}}]{Zioutas:2024cmy}%
  \BibitemOpen
  \bibfield  {author} {\bibinfo {author} {\bibfnamefont {K.}~\bibnamefont
  {Zioutas}} \emph {et~al.},\ }\href@noop {} {\  (\bibinfo {year}
  {2024}{\natexlab{a}})},\ \Eprint {http://arxiv.org/abs/2403.05608}
  {arXiv:2403.05608 [hep-ex]} \BibitemShut {NoStop}%
\bibitem [{\citenamefont {Adair}\ \emph {et~al.}(2024)\citenamefont {Adair}
  \emph {et~al.}}]{Adair:2024wki}%
  \BibitemOpen
  \bibfield  {author} {\bibinfo {author} {\bibfnamefont {C.~M.}\ \bibnamefont
  {Adair}} \emph {et~al.},\ }\href@noop {} {\  (\bibinfo {year} {2024})},\
  \Eprint {http://arxiv.org/abs/2405.10972} {arXiv:2405.10972 [hep-ex]}
  \BibitemShut {NoStop}%
\bibitem [{\citenamefont {Freese}\ \emph {et~al.}(2013)\citenamefont {Freese},
  \citenamefont {Lisanti},\ and\ \citenamefont {Savage}}]{Freese:2012xd}%
  \BibitemOpen
  \bibfield  {author} {\bibinfo {author} {\bibfnamefont {K.}~\bibnamefont
  {Freese}}, \bibinfo {author} {\bibfnamefont {M.}~\bibnamefont {Lisanti}}, \
  and\ \bibinfo {author} {\bibfnamefont {C.}~\bibnamefont {Savage}},\ }\href
  {\doibase 10.1103/RevModPhys.85.1561} {\bibfield  {journal} {\bibinfo
  {journal} {Rev. Mod. Phys.}\ }\textbf {\bibinfo {volume} {85}},\ \bibinfo
  {pages} {1561} (\bibinfo {year} {2013})},\ \Eprint
  {http://arxiv.org/abs/1209.3339} {arXiv:1209.3339 [astro-ph.CO]} \BibitemShut
  {NoStop}%
\bibitem [{\citenamefont {Zioutas}\ \emph
  {et~al.}(2024{\natexlab{b}})\citenamefont {Zioutas}, \citenamefont
  {Anastassopoulos}, \citenamefont {Argiriou}, \citenamefont {Cantatore},
  \citenamefont {Cetin}, \citenamefont {Gardikiotis}, \citenamefont {Guo},
  \citenamefont {Haralambous}, \citenamefont {Hoffmann}, \citenamefont
  {Hofmann}, \citenamefont {Karuza}, \citenamefont {Kryemadhi}, \citenamefont
  {Maroudas}, \citenamefont {Mastronikolis}, \citenamefont {Oikonomou},
  \citenamefont {Ozbozduman},\ and\ \citenamefont
  {Semertzidis}}]{Zioutas_geoscience}%
  \BibitemOpen
  \bibfield  {author} {\bibinfo {author} {\bibfnamefont {K.}~\bibnamefont
  {Zioutas}}, \bibinfo {author} {\bibfnamefont {V.}~\bibnamefont
  {Anastassopoulos}}, \bibinfo {author} {\bibfnamefont {A.}~\bibnamefont
  {Argiriou}}, \bibinfo {author} {\bibfnamefont {G.}~\bibnamefont {Cantatore}},
  \bibinfo {author} {\bibfnamefont {S.}~\bibnamefont {Cetin}}, \bibinfo
  {author} {\bibfnamefont {A.}~\bibnamefont {Gardikiotis}}, \bibinfo {author}
  {\bibfnamefont {J.}~\bibnamefont {Guo}}, \bibinfo {author} {\bibfnamefont
  {H.}~\bibnamefont {Haralambous}}, \bibinfo {author} {\bibfnamefont
  {D.}~\bibnamefont {Hoffmann}}, \bibinfo {author} {\bibfnamefont
  {S.}~\bibnamefont {Hofmann}}, \bibinfo {author} {\bibfnamefont
  {M.}~\bibnamefont {Karuza}}, \bibinfo {author} {\bibfnamefont
  {A.}~\bibnamefont {Kryemadhi}}, \bibinfo {author} {\bibfnamefont
  {M.}~\bibnamefont {Maroudas}}, \bibinfo {author} {\bibfnamefont
  {A.}~\bibnamefont {Mastronikolis}}, \bibinfo {author} {\bibfnamefont
  {C.}~\bibnamefont {Oikonomou}}, \bibinfo {author} {\bibfnamefont
  {K.}~\bibnamefont {Ozbozduman}}, \ and\ \bibinfo {author} {\bibfnamefont
  {Y.}~\bibnamefont {Semertzidis}},\ }in\ \href@noop {} {\emph {\bibinfo
  {booktitle} {Recent Research on Geotechnical Engineering, Remote Sensing,
  Geophysics and Earthquake Seismology}}},\ \bibinfo {editor} {edited by\
  \bibinfo {editor} {\bibfnamefont {A.}~\bibnamefont {{\c{C}}iner}}, \bibinfo
  {editor} {\bibfnamefont {Z.~A.}\ \bibnamefont {Erg{\"u}ler}}, \bibinfo
  {editor} {\bibfnamefont {M.}~\bibnamefont {Bezzeghoud}}, \bibinfo {editor}
  {\bibfnamefont {M.}~\bibnamefont {Ustuner}}, \bibinfo {editor} {\bibfnamefont
  {M.}~\bibnamefont {Eshagh}}, \bibinfo {editor} {\bibfnamefont
  {H.}~\bibnamefont {El-Askary}}, \bibinfo {editor} {\bibfnamefont
  {A.}~\bibnamefont {Biswas}}, \bibinfo {editor} {\bibfnamefont
  {L.}~\bibnamefont {Gasperini}}, \bibinfo {editor} {\bibfnamefont {K.-G.}\
  \bibnamefont {Hinzen}}, \bibinfo {editor} {\bibfnamefont {M.}~\bibnamefont
  {Karakus}}, \bibinfo {editor} {\bibfnamefont {C.}~\bibnamefont {Comina}},
  \bibinfo {editor} {\bibfnamefont {A.}~\bibnamefont {Karrech}}, \bibinfo
  {editor} {\bibfnamefont {A.}~\bibnamefont {Polonia}}, \ and\ \bibinfo
  {editor} {\bibfnamefont {H.~I.}\ \bibnamefont {Chamin{\'e}}}}\ (\bibinfo
  {publisher} {Springer Nature Switzerland},\ \bibinfo {address} {Cham},\
  \bibinfo {year} {2024})\ pp.\ \bibinfo {pages} {415--419}\BibitemShut
  {NoStop}%
\bibitem [{\citenamefont {Tulin}\ and\ \citenamefont
  {Yu}(2018)}]{Tulin:2017ara}%
  \BibitemOpen
  \bibfield  {author} {\bibinfo {author} {\bibfnamefont {S.}~\bibnamefont
  {Tulin}}\ and\ \bibinfo {author} {\bibfnamefont {H.-B.}\ \bibnamefont {Yu}},\
  }\href {\doibase 10.1016/j.physrep.2017.11.004} {\bibfield  {journal}
  {\bibinfo  {journal} {Phys. Rept.}\ }\textbf {\bibinfo {volume} {730}},\
  \bibinfo {pages} {1} (\bibinfo {year} {2018})},\ \Eprint
  {http://arxiv.org/abs/1705.02358} {arXiv:1705.02358 [hep-ph]} \BibitemShut
  {NoStop}%
\bibitem [{\citenamefont {{Witten}}(1984)}]{Witten:1984rs}%
  \BibitemOpen
  \bibfield  {author} {\bibinfo {author} {\bibfnamefont {E.}~\bibnamefont
  {{Witten}}},\ }\href {\doibase 10.1103/PhysRevD.30.272} {\bibfield  {journal}
  {\bibinfo  {journal} {\prd}\ }\textbf {\bibinfo {volume} {30}},\ \bibinfo
  {pages} {272} (\bibinfo {year} {1984})}\BibitemShut {NoStop}%
\bibitem [{\citenamefont {{Farhi}}\ and\ \citenamefont
  {{Jaffe}}(1984)}]{Farhi:1984qu}%
  \BibitemOpen
  \bibfield  {author} {\bibinfo {author} {\bibfnamefont {E.}~\bibnamefont
  {{Farhi}}}\ and\ \bibinfo {author} {\bibfnamefont {R.~L.}\ \bibnamefont
  {{Jaffe}}},\ }\href {\doibase 10.1103/PhysRevD.30.2379} {\bibfield  {journal}
  {\bibinfo  {journal} {\prd}\ }\textbf {\bibinfo {volume} {30}},\ \bibinfo
  {pages} {2379} (\bibinfo {year} {1984})}\BibitemShut {NoStop}%
\bibitem [{\citenamefont {{De Rujula}}\ and\ \citenamefont
  {{Glashow}}(1984)}]{DeRujula:1984axn}%
  \BibitemOpen
  \bibfield  {author} {\bibinfo {author} {\bibfnamefont {A.}~\bibnamefont {{De
  Rujula}}}\ and\ \bibinfo {author} {\bibfnamefont {S.~L.}\ \bibnamefont
  {{Glashow}}},\ }\href {\doibase 10.1038/312734a0} {\bibfield  {journal}
  {\bibinfo  {journal} {\nat}\ }\textbf {\bibinfo {volume} {312}},\ \bibinfo
  {pages} {734} (\bibinfo {year} {1984})}\BibitemShut {NoStop}%
\bibitem [{\citenamefont {Abbasi}\ \emph {et~al.}(2017)\citenamefont {Abbasi}
  \emph {et~al.}}]{Abbasi:2017rvx}%
  \BibitemOpen
  \bibfield  {author} {\bibinfo {author} {\bibfnamefont {R.}~\bibnamefont
  {Abbasi}} \emph {et~al.} (\bibinfo {collaboration} {Telescope Array
  Project}),\ }\href {\doibase 10.1016/j.physleta.2017.06.022} {\bibfield
  {journal} {\bibinfo  {journal} {Phys. Lett. A}\ }\textbf {\bibinfo {volume}
  {381}},\ \bibinfo {pages} {2565} (\bibinfo {year} {2017})}\BibitemShut
  {NoStop}%
\bibitem [{\citenamefont {Okuda}(2019)}]{Okuda_2019}%
  \BibitemOpen
  \bibfield  {author} {\bibinfo {author} {\bibfnamefont {T.}~\bibnamefont
  {Okuda}},\ }\href {\doibase 10.1088/1742-6596/1181/1/012067} {\bibfield
  {journal} {\bibinfo  {journal} {Journal of Physics: Conference Series}\
  }\textbf {\bibinfo {volume} {1181}},\ \bibinfo {pages} {012067} (\bibinfo
  {year} {2019})}\BibitemShut {NoStop}%
\bibitem [{\citenamefont
  {Zhitnitsky}(2021{\natexlab{b}})}]{Zhitnitsky:2020shd}%
  \BibitemOpen
  \bibfield  {author} {\bibinfo {author} {\bibfnamefont {A.}~\bibnamefont
  {Zhitnitsky}},\ }\href {\doibase 10.1088/1361-6471/abd457} {\bibfield
  {journal} {\bibinfo  {journal} {J. Phys. G}\ }\textbf {\bibinfo {volume}
  {48}},\ \bibinfo {pages} {065201} (\bibinfo {year} {2021}{\natexlab{b}})},\
  \Eprint {http://arxiv.org/abs/2008.04325} {arXiv:2008.04325 [hep-ph]}
  \BibitemShut {NoStop}%
\bibitem [{\citenamefont {Abreu}\ \emph {et~al.}(2021)\citenamefont {Abreu}
  \emph {et~al.}}]{PierreAuger:2021int}%
  \BibitemOpen
  \bibfield  {author} {\bibinfo {author} {\bibfnamefont {P.}~\bibnamefont
  {Abreu}} \emph {et~al.} (\bibinfo {collaboration} {Pierre Auger}),\ }\href
  {\doibase 10.22323/1.395.0395} {\bibfield  {journal} {\bibinfo  {journal}
  {PoS}\ }\textbf {\bibinfo {volume} {ICRC2021}},\ \bibinfo {pages} {395}
  (\bibinfo {year} {2021})}\BibitemShut {NoStop}%
\bibitem [{\citenamefont {{Colalillo}}(2019)}]{2019EPJWC.19703003C}%
  \BibitemOpen
  \bibfield  {author} {\bibinfo {author} {\bibfnamefont {R.}~\bibnamefont
  {{Colalillo}}},\ }in\ \href {\doibase 10.1051/epjconf/201919703003} {\emph
  {\bibinfo {booktitle} {European Physical Journal Web of Conferences}}},\
  \bibinfo {series} {European Physical Journal Web of Conferences}, Vol.\
  \bibinfo {volume} {197}\ (\bibinfo {year} {2019})\ p.\ \bibinfo {pages}
  {03003}\BibitemShut {NoStop}%
\bibitem [{\citenamefont {Colalillo}(2017)}]{Colalillo:2017uC}%
  \BibitemOpen
  \bibfield  {author} {\bibinfo {author} {\bibfnamefont {R.}~\bibnamefont
  {Colalillo}},\ }\href {\doibase 10.22323/1.301.0314} {\bibfield  {journal}
  {\bibinfo  {journal} {PoS}\ }\textbf {\bibinfo {volume} {ICRC2017}},\
  \bibinfo {pages} {314} (\bibinfo {year} {2017})}\BibitemShut {NoStop}%
\bibitem [{\citenamefont
  {Zhitnitsky}(2022{\natexlab{a}})}]{Zhitnitsky:2022swb}%
  \BibitemOpen
  \bibfield  {author} {\bibinfo {author} {\bibfnamefont {A.}~\bibnamefont
  {Zhitnitsky}},\ }\href {\doibase 10.1088/1361-6471/ac8569} {\bibfield
  {journal} {\bibinfo  {journal} {J. Phys. G}\ }\textbf {\bibinfo {volume}
  {49}},\ \bibinfo {pages} {105201} (\bibinfo {year} {2022}{\natexlab{a}})},\
  \Eprint {http://arxiv.org/abs/2203.08160} {arXiv:2203.08160 [hep-ph]}
  \BibitemShut {NoStop}%
\bibitem [{\citenamefont {Gorham}\ \emph {et~al.}(2016)\citenamefont {Gorham}
  \emph {et~al.}}]{Gorham:2016zah}%
  \BibitemOpen
  \bibfield  {author} {\bibinfo {author} {\bibfnamefont {P.~W.}\ \bibnamefont
  {Gorham}} \emph {et~al.} (\bibinfo {collaboration} {ANITA}),\ }\href
  {\doibase 10.1103/PhysRevLett.117.071101} {\bibfield  {journal} {\bibinfo
  {journal} {Phys. Rev. Lett.}\ }\textbf {\bibinfo {volume} {117}},\ \bibinfo
  {pages} {071101} (\bibinfo {year} {2016})},\ \Eprint
  {http://arxiv.org/abs/1603.05218} {arXiv:1603.05218 [astro-ph.HE]}
  \BibitemShut {NoStop}%
\bibitem [{\citenamefont {Gorham}\ \emph {et~al.}(2018)\citenamefont {Gorham}
  \emph {et~al.}}]{Gorham:2018ydl}%
  \BibitemOpen
  \bibfield  {author} {\bibinfo {author} {\bibfnamefont {P.~W.}\ \bibnamefont
  {Gorham}} \emph {et~al.} (\bibinfo {collaboration} {ANITA}),\ }\href
  {\doibase 10.1103/PhysRevLett.121.161102} {\bibfield  {journal} {\bibinfo
  {journal} {Phys. Rev. Lett.}\ }\textbf {\bibinfo {volume} {121}},\ \bibinfo
  {pages} {161102} (\bibinfo {year} {2018})},\ \Eprint
  {http://arxiv.org/abs/1803.05088} {arXiv:1803.05088 [astro-ph.HE]}
  \BibitemShut {NoStop}%
\bibitem [{\citenamefont {Liang}\ and\ \citenamefont
  {Zhitnitsky}(2022)}]{Liang:2021rnv}%
  \BibitemOpen
  \bibfield  {author} {\bibinfo {author} {\bibfnamefont {X.}~\bibnamefont
  {Liang}}\ and\ \bibinfo {author} {\bibfnamefont {A.}~\bibnamefont
  {Zhitnitsky}},\ }\href {\doibase 10.1103/PhysRevD.106.063022} {\bibfield
  {journal} {\bibinfo  {journal} {Phys. Rev. D}\ }\textbf {\bibinfo {volume}
  {106}},\ \bibinfo {pages} {063022} (\bibinfo {year} {2022})},\ \Eprint
  {http://arxiv.org/abs/2105.01668} {arXiv:2105.01668 [hep-ph]} \BibitemShut
  {NoStop}%
\bibitem [{\citenamefont {{Beznosko}}\ \emph {et~al.}(2017)\citenamefont
  {{Beznosko}}, \citenamefont {{Beisembaev}}, \citenamefont {{Baigarin}},
  \citenamefont {{Beisembaeva}}, \citenamefont {{Dalkarov}}, \citenamefont
  {{Ryabov}}, \citenamefont {{Sadykov}}, \citenamefont {{Shaulov}},
  \citenamefont {{Stepanov}}, \citenamefont {{Vildanova}}, \citenamefont
  {{Vildanov}},\ and\ \citenamefont {{Zhukov}}}]{2017EPJWC.14514001B}%
  \BibitemOpen
  \bibfield  {author} {\bibinfo {author} {\bibfnamefont {D.}~\bibnamefont
  {{Beznosko}}}, \bibinfo {author} {\bibfnamefont {R.}~\bibnamefont
  {{Beisembaev}}}, \bibinfo {author} {\bibfnamefont {K.}~\bibnamefont
  {{Baigarin}}}, \bibinfo {author} {\bibfnamefont {E.}~\bibnamefont
  {{Beisembaeva}}}, \bibinfo {author} {\bibfnamefont {O.}~\bibnamefont
  {{Dalkarov}}}, \bibinfo {author} {\bibfnamefont {V.}~\bibnamefont
  {{Ryabov}}}, \bibinfo {author} {\bibfnamefont {T.}~\bibnamefont {{Sadykov}}},
  \bibinfo {author} {\bibfnamefont {S.}~\bibnamefont {{Shaulov}}}, \bibinfo
  {author} {\bibfnamefont {A.}~\bibnamefont {{Stepanov}}}, \bibinfo {author}
  {\bibfnamefont {M.}~\bibnamefont {{Vildanova}}}, \bibinfo {author}
  {\bibfnamefont {N.}~\bibnamefont {{Vildanov}}}, \ and\ \bibinfo {author}
  {\bibfnamefont {V.}~\bibnamefont {{Zhukov}}},\ }in\ \href {\doibase
  10.1051/epjconf/201714514001} {\emph {\bibinfo {booktitle} {European Physical
  Journal Web of Conferences}}},\ \bibinfo {series} {European Physical Journal
  Web of Conferences}, Vol.\ \bibinfo {volume} {145}\ (\bibinfo {year} {2017})\
  p.\ \bibinfo {pages} {14001}\BibitemShut {NoStop}%
\bibitem [{\citenamefont {Beznosko}\ \emph {et~al.}(2019)\citenamefont
  {Beznosko}, \citenamefont {Beisembaev}, \citenamefont {Beisembaeva},
  \citenamefont {Dalkarov}, \citenamefont {Mossunov}, \citenamefont {Ryabov},
  \citenamefont {Shaulov}, \citenamefont {Vildanova}, \citenamefont {Zhukov},
  \citenamefont {Baigarin},\ and\ \citenamefont {Sadykov}}]{Beznosko:2019cI}%
  \BibitemOpen
  \bibfield  {author} {\bibinfo {author} {\bibfnamefont {D.}~\bibnamefont
  {Beznosko}}, \bibinfo {author} {\bibfnamefont {R.}~\bibnamefont
  {Beisembaev}}, \bibinfo {author} {\bibfnamefont {E.}~\bibnamefont
  {Beisembaeva}}, \bibinfo {author} {\bibfnamefont {O.~D.}\ \bibnamefont
  {Dalkarov}}, \bibinfo {author} {\bibfnamefont {V.}~\bibnamefont {Mossunov}},
  \bibinfo {author} {\bibfnamefont {V.}~\bibnamefont {Ryabov}}, \bibinfo
  {author} {\bibfnamefont {S.}~\bibnamefont {Shaulov}}, \bibinfo {author}
  {\bibfnamefont {M.}~\bibnamefont {Vildanova}}, \bibinfo {author}
  {\bibfnamefont {V.}~\bibnamefont {Zhukov}}, \bibinfo {author} {\bibfnamefont
  {K.}~\bibnamefont {Baigarin}}, \ and\ \bibinfo {author} {\bibfnamefont
  {T.}~\bibnamefont {Sadykov}},\ }\href {\doibase 10.22323/1.358.0195}
  {\bibfield  {journal} {\bibinfo  {journal} {PoS}\ }\textbf {\bibinfo {volume}
  {ICRC2019}},\ \bibinfo {pages} {195} (\bibinfo {year} {2019})}\BibitemShut
  {NoStop}%
\bibitem [{\citenamefont
  {Zhitnitsky}(2021{\natexlab{c}})}]{Zhitnitsky:2021qhj}%
  \BibitemOpen
  \bibfield  {author} {\bibinfo {author} {\bibfnamefont {A.}~\bibnamefont
  {Zhitnitsky}},\ }\href {\doibase 10.3390/universe7100384} {\bibfield
  {journal} {\bibinfo  {journal} {Universe}\ }\textbf {\bibinfo {volume} {7}},\
  \bibinfo {pages} {384} (\bibinfo {year} {2021}{\natexlab{c}})},\ \Eprint
  {http://arxiv.org/abs/2108.04826} {arXiv:2108.04826 [hep-ph]} \BibitemShut
  {NoStop}%
\bibitem [{\citenamefont {Budker}\ \emph {et~al.}(2022)\citenamefont {Budker},
  \citenamefont {Flambaum},\ and\ \citenamefont {Zhitnitsky}}]{Budker:2020mqk}%
  \BibitemOpen
  \bibfield  {author} {\bibinfo {author} {\bibfnamefont {D.}~\bibnamefont
  {Budker}}, \bibinfo {author} {\bibfnamefont {V.~V.}\ \bibnamefont
  {Flambaum}}, \ and\ \bibinfo {author} {\bibfnamefont {A.}~\bibnamefont
  {Zhitnitsky}},\ }\href {\doibase 10.3390/sym14030459} {\bibfield  {journal}
  {\bibinfo  {journal} {Symmetry}\ }\textbf {\bibinfo {volume} {14}},\ \bibinfo
  {pages} {459} (\bibinfo {year} {2022})},\ \Eprint
  {http://arxiv.org/abs/2003.07363} {arXiv:2003.07363 [hep-ph]} \BibitemShut
  {NoStop}%
\bibitem [{\citenamefont {Lawson}\ \emph {et~al.}(2019)\citenamefont {Lawson},
  \citenamefont {Liang}, \citenamefont {Mead}, \citenamefont {Siddiqui},
  \citenamefont {Van~Waerbeke},\ and\ \citenamefont
  {Zhitnitsky}}]{Lawson:2019cvy}%
  \BibitemOpen
  \bibfield  {author} {\bibinfo {author} {\bibfnamefont {K.}~\bibnamefont
  {Lawson}}, \bibinfo {author} {\bibfnamefont {X.}~\bibnamefont {Liang}},
  \bibinfo {author} {\bibfnamefont {A.}~\bibnamefont {Mead}}, \bibinfo {author}
  {\bibfnamefont {M.~S.~R.}\ \bibnamefont {Siddiqui}}, \bibinfo {author}
  {\bibfnamefont {L.}~\bibnamefont {Van~Waerbeke}}, \ and\ \bibinfo {author}
  {\bibfnamefont {A.}~\bibnamefont {Zhitnitsky}},\ }\href {\doibase
  10.1103/PhysRevD.100.043531} {\bibfield  {journal} {\bibinfo  {journal}
  {Phys. Rev. D}\ }\textbf {\bibinfo {volume} {100}},\ \bibinfo {pages}
  {043531} (\bibinfo {year} {2019})},\ \Eprint
  {http://arxiv.org/abs/1905.00022} {arXiv:1905.00022 [astro-ph.CO]}
  \BibitemShut {NoStop}%
\bibitem [{\citenamefont {{Liang}}\ and\ \citenamefont
  {{Zhitnitsky}}(2016)}]{Liang:2016tqc}%
  \BibitemOpen
  \bibfield  {author} {\bibinfo {author} {\bibfnamefont {X.}~\bibnamefont
  {{Liang}}}\ and\ \bibinfo {author} {\bibfnamefont {A.}~\bibnamefont
  {{Zhitnitsky}}},\ }\href {\doibase 10.1103/PhysRevD.94.083502} {\bibfield
  {journal} {\bibinfo  {journal} {\prd}\ }\textbf {\bibinfo {volume} {94}},\
  \bibinfo {eid} {083502} (\bibinfo {year} {2016})},\ \Eprint
  {http://arxiv.org/abs/1606.00435} {arXiv:1606.00435 [hep-ph]} \BibitemShut
  {NoStop}%
\bibitem [{\citenamefont {{Ge}}\ \emph {et~al.}(2017)\citenamefont {{Ge}},
  \citenamefont {{Liang}},\ and\ \citenamefont {{Zhitnitsky}}}]{Ge:2017ttc}%
  \BibitemOpen
  \bibfield  {author} {\bibinfo {author} {\bibfnamefont {S.}~\bibnamefont
  {{Ge}}}, \bibinfo {author} {\bibfnamefont {X.}~\bibnamefont {{Liang}}}, \
  and\ \bibinfo {author} {\bibfnamefont {A.}~\bibnamefont {{Zhitnitsky}}},\
  }\href {\doibase 10.1103/PhysRevD.96.063514} {\bibfield  {journal} {\bibinfo
  {journal} {\prd}\ }\textbf {\bibinfo {volume} {96}},\ \bibinfo {eid} {063514}
  (\bibinfo {year} {2017})},\ \Eprint {http://arxiv.org/abs/1702.04354}
  {arXiv:1702.04354 [hep-ph]} \BibitemShut {NoStop}%
\bibitem [{\citenamefont {{Ge}}\ \emph {et~al.}(2018)\citenamefont {{Ge}},
  \citenamefont {{Liang}},\ and\ \citenamefont {{Zhitnitsky}}}]{Ge:2017idw}%
  \BibitemOpen
  \bibfield  {author} {\bibinfo {author} {\bibfnamefont {S.}~\bibnamefont
  {{Ge}}}, \bibinfo {author} {\bibfnamefont {X.}~\bibnamefont {{Liang}}}, \
  and\ \bibinfo {author} {\bibfnamefont {A.}~\bibnamefont {{Zhitnitsky}}},\
  }\href {\doibase 10.1103/PhysRevD.97.043008} {\bibfield  {journal} {\bibinfo
  {journal} {\prd}\ }\textbf {\bibinfo {volume} {97}},\ \bibinfo {eid} {043008}
  (\bibinfo {year} {2018})},\ \Eprint {http://arxiv.org/abs/1711.06271}
  {arXiv:1711.06271 [hep-ph]} \BibitemShut {NoStop}%
\bibitem [{\citenamefont {Ge}\ \emph {et~al.}(2019)\citenamefont {Ge},
  \citenamefont {Lawson},\ and\ \citenamefont {Zhitnitsky}}]{Ge:2019voa}%
  \BibitemOpen
  \bibfield  {author} {\bibinfo {author} {\bibfnamefont {S.}~\bibnamefont
  {Ge}}, \bibinfo {author} {\bibfnamefont {K.}~\bibnamefont {Lawson}}, \ and\
  \bibinfo {author} {\bibfnamefont {A.}~\bibnamefont {Zhitnitsky}},\ }\href
  {\doibase 10.1103/PhysRevD.99.116017} {\bibfield  {journal} {\bibinfo
  {journal} {Phys. Rev. D}\ }\textbf {\bibinfo {volume} {99}},\ \bibinfo
  {pages} {116017} (\bibinfo {year} {2019})},\ \Eprint
  {http://arxiv.org/abs/1903.05090} {arXiv:1903.05090 [hep-ph]} \BibitemShut
  {NoStop}%
\bibitem [{\citenamefont {{Zhitnitsky}}(2006)}]{Zhitnitsky:2006vt}%
  \BibitemOpen
  \bibfield  {author} {\bibinfo {author} {\bibfnamefont {A.}~\bibnamefont
  {{Zhitnitsky}}},\ }\href {\doibase 10.1103/PhysRevD.74.043515} {\bibfield
  {journal} {\bibinfo  {journal} {\prd}\ }\textbf {\bibinfo {volume} {74}},\
  \bibinfo {eid} {043515} (\bibinfo {year} {2006})},\ \Eprint
  {http://arxiv.org/abs/astro-ph/0603064} {astro-ph/0603064} \BibitemShut
  {NoStop}%
\bibitem [{\citenamefont {Flambaum}\ and\ \citenamefont
  {Zhitnitsky}(2019)}]{Flambaum:2018ohm}%
  \BibitemOpen
  \bibfield  {author} {\bibinfo {author} {\bibfnamefont {V.~V.}\ \bibnamefont
  {Flambaum}}\ and\ \bibinfo {author} {\bibfnamefont {A.~R.}\ \bibnamefont
  {Zhitnitsky}},\ }\href {\doibase 10.1103/PhysRevD.99.023517} {\bibfield
  {journal} {\bibinfo  {journal} {Phys. Rev. D}\ }\textbf {\bibinfo {volume}
  {99}},\ \bibinfo {pages} {023517} (\bibinfo {year} {2019})},\ \Eprint
  {http://arxiv.org/abs/1811.01965} {arXiv:1811.01965 [hep-ph]} \BibitemShut
  {NoStop}%
\bibitem [{\citenamefont {Singh~Sidhu}\ \emph {et~al.}(2020)\citenamefont
  {Singh~Sidhu}, \citenamefont {Scherrer},\ and\ \citenamefont
  {Starkman}}]{SinghSidhu:2020cxw}%
  \BibitemOpen
  \bibfield  {author} {\bibinfo {author} {\bibfnamefont {J.}~\bibnamefont
  {Singh~Sidhu}}, \bibinfo {author} {\bibfnamefont {R.~J.}\ \bibnamefont
  {Scherrer}}, \ and\ \bibinfo {author} {\bibfnamefont {G.}~\bibnamefont
  {Starkman}},\ }\href {\doibase 10.1016/j.physletb.2020.135574} {\bibfield
  {journal} {\bibinfo  {journal} {Phys. Lett. B}\ }\textbf {\bibinfo {volume}
  {807}},\ \bibinfo {pages} {135574} (\bibinfo {year} {2020})},\ \Eprint
  {http://arxiv.org/abs/2006.01200} {arXiv:2006.01200 [astro-ph.CO]}
  \BibitemShut {NoStop}%
\bibitem [{\citenamefont {Santill\'an}\ and\ \citenamefont
  {Morano}(2021)}]{Santillan:2020lbj}%
  \BibitemOpen
  \bibfield  {author} {\bibinfo {author} {\bibfnamefont {O.~P.}\ \bibnamefont
  {Santill\'an}}\ and\ \bibinfo {author} {\bibfnamefont {A.}~\bibnamefont
  {Morano}},\ }\href {\doibase 10.1103/PhysRevD.104.083530} {\bibfield
  {journal} {\bibinfo  {journal} {Phys. Rev. D}\ }\textbf {\bibinfo {volume}
  {104}},\ \bibinfo {pages} {083530} (\bibinfo {year} {2021})},\ \Eprint
  {http://arxiv.org/abs/2011.06747} {arXiv:2011.06747 [hep-ph]} \BibitemShut
  {NoStop}%
\bibitem [{\citenamefont {Lawson}\ and\ \citenamefont
  {Zhitnitsky}(2019)}]{Lawson:2018qkc}%
  \BibitemOpen
  \bibfield  {author} {\bibinfo {author} {\bibfnamefont {K.}~\bibnamefont
  {Lawson}}\ and\ \bibinfo {author} {\bibfnamefont {A.~R.}\ \bibnamefont
  {Zhitnitsky}},\ }\href {\doibase 10.1016/j.dark.2019.100295} {\bibfield
  {journal} {\bibinfo  {journal} {Phys. Dark Univ.}\ }\textbf {\bibinfo
  {volume} {24}},\ \bibinfo {pages} {100295} (\bibinfo {year} {2019})},\
  \Eprint {http://arxiv.org/abs/1804.07340} {arXiv:1804.07340 [hep-ph]}
  \BibitemShut {NoStop}%
\bibitem [{\citenamefont {Jacobs}\ \emph {et~al.}(2015)\citenamefont {Jacobs},
  \citenamefont {Starkman},\ and\ \citenamefont {Lynn}}]{Jacobs:2014yca}%
  \BibitemOpen
  \bibfield  {author} {\bibinfo {author} {\bibfnamefont {D.~M.}\ \bibnamefont
  {Jacobs}}, \bibinfo {author} {\bibfnamefont {G.~D.}\ \bibnamefont
  {Starkman}}, \ and\ \bibinfo {author} {\bibfnamefont {B.~W.}\ \bibnamefont
  {Lynn}},\ }\href {\doibase 10.1093/mnras/stv774} {\bibfield  {journal}
  {\bibinfo  {journal} {Mon. Not. Roy. Astron. Soc.}\ }\textbf {\bibinfo
  {volume} {450}},\ \bibinfo {pages} {3418} (\bibinfo {year} {2015})},\ \Eprint
  {http://arxiv.org/abs/1410.2236} {arXiv:1410.2236 [astro-ph.CO]} \BibitemShut
  {NoStop}%
\bibitem [{\citenamefont {Alford}\ \emph {et~al.}(2008)\citenamefont {Alford},
  \citenamefont {Schmitt}, \citenamefont {Rajagopal},\ and\ \citenamefont
  {Sch\"afer}}]{Alford:2007xm}%
  \BibitemOpen
  \bibfield  {author} {\bibinfo {author} {\bibfnamefont {M.~G.}\ \bibnamefont
  {Alford}}, \bibinfo {author} {\bibfnamefont {A.}~\bibnamefont {Schmitt}},
  \bibinfo {author} {\bibfnamefont {K.}~\bibnamefont {Rajagopal}}, \ and\
  \bibinfo {author} {\bibfnamefont {T.}~\bibnamefont {Sch\"afer}},\ }\href
  {\doibase 10.1103/RevModPhys.80.1455} {\bibfield  {journal} {\bibinfo
  {journal} {Rev. Mod. Phys.}\ }\textbf {\bibinfo {volume} {80}},\ \bibinfo
  {pages} {1455} (\bibinfo {year} {2008})},\ \Eprint
  {http://arxiv.org/abs/0709.4635} {arXiv:0709.4635 [hep-ph]} \BibitemShut
  {NoStop}%
\bibitem [{\citenamefont {Ge}\ \emph {et~al.}(2020)\citenamefont {Ge},
  \citenamefont {Siddiqui}, \citenamefont {Van~Waerbeke},\ and\ \citenamefont
  {Zhitnitsky}}]{Ge:2020xvf}%
  \BibitemOpen
  \bibfield  {author} {\bibinfo {author} {\bibfnamefont {S.}~\bibnamefont
  {Ge}}, \bibinfo {author} {\bibfnamefont {M.~S.~R.}\ \bibnamefont {Siddiqui}},
  \bibinfo {author} {\bibfnamefont {L.}~\bibnamefont {Van~Waerbeke}}, \ and\
  \bibinfo {author} {\bibfnamefont {A.}~\bibnamefont {Zhitnitsky}},\ }\href
  {\doibase 10.1103/PhysRevD.102.123021} {\bibfield  {journal} {\bibinfo
  {journal} {Phys. Rev. D}\ }\textbf {\bibinfo {volume} {102}},\ \bibinfo
  {pages} {123021} (\bibinfo {year} {2020})},\ \Eprint
  {http://arxiv.org/abs/2009.00004} {arXiv:2009.00004 [astro-ph.HE]}
  \BibitemShut {NoStop}%
\bibitem [{\citenamefont {{Forbes}}\ and\ \citenamefont
  {{Zhitnitsky}}(2008)}]{Forbes:2008uf}%
  \BibitemOpen
  \bibfield  {author} {\bibinfo {author} {\bibfnamefont {M.~M.}\ \bibnamefont
  {{Forbes}}}\ and\ \bibinfo {author} {\bibfnamefont {A.~R.}\ \bibnamefont
  {{Zhitnitsky}}},\ }\href {\doibase 10.1103/PhysRevD.78.083505} {\bibfield
  {journal} {\bibinfo  {journal} {\prd}\ }\textbf {\bibinfo {volume} {78}},\
  \bibinfo {eid} {083505} (\bibinfo {year} {2008})},\ \Eprint
  {http://arxiv.org/abs/0802.3830} {arXiv:0802.3830} \BibitemShut {NoStop}%
\bibitem [{\citenamefont {Judge}\ \emph {et~al.}(2014)\citenamefont {Judge},
  \citenamefont {Kleint}, \citenamefont {Donea}, \citenamefont {Dalda},\ and\
  \citenamefont {Fletcher}}]{Judge_2014}%
  \BibitemOpen
  \bibfield  {author} {\bibinfo {author} {\bibfnamefont {P.~G.}\ \bibnamefont
  {Judge}}, \bibinfo {author} {\bibfnamefont {L.}~\bibnamefont {Kleint}},
  \bibinfo {author} {\bibfnamefont {A.}~\bibnamefont {Donea}}, \bibinfo
  {author} {\bibfnamefont {A.~S.}\ \bibnamefont {Dalda}}, \ and\ \bibinfo
  {author} {\bibfnamefont {L.}~\bibnamefont {Fletcher}},\ }\href {\doibase
  10.1088/0004-637x/796/2/85} {\bibfield  {journal} {\bibinfo  {journal} {The
  Astrophysical Journal}\ }\textbf {\bibinfo {volume} {796}},\ \bibinfo {pages}
  {85} (\bibinfo {year} {2014})}\BibitemShut {NoStop}%
\bibitem [{\citenamefont {{Matthews}}\ \emph {et~al.}(2015)\citenamefont
  {{Matthews}}, \citenamefont {{Harra}}, \citenamefont {{Zharkov}},\ and\
  \citenamefont {{Green}}}]{2015ApJ...812...35M}%
  \BibitemOpen
  \bibfield  {author} {\bibinfo {author} {\bibfnamefont {S.~A.}\ \bibnamefont
  {{Matthews}}}, \bibinfo {author} {\bibfnamefont {L.~K.}\ \bibnamefont
  {{Harra}}}, \bibinfo {author} {\bibfnamefont {S.}~\bibnamefont {{Zharkov}}},
  \ and\ \bibinfo {author} {\bibfnamefont {L.~M.}\ \bibnamefont {{Green}}},\
  }\href {\doibase 10.1088/0004-637X/812/1/35} {\bibfield  {journal} {\bibinfo
  {journal} {\apj}\ }\textbf {\bibinfo {volume} {812}},\ \bibinfo {eid} {35}
  (\bibinfo {year} {2015})},\ \Eprint {http://arxiv.org/abs/1508.07216}
  {arXiv:1508.07216 [astro-ph.SR]} \BibitemShut {NoStop}%
\bibitem [{\citenamefont {Zhitnitsky}(2018)}]{Zhitnitsky:2018mav}%
  \BibitemOpen
  \bibfield  {author} {\bibinfo {author} {\bibfnamefont {A.}~\bibnamefont
  {Zhitnitsky}},\ }\href {\doibase 10.1016/j.dark.2018.08.001} {\bibfield
  {journal} {\bibinfo  {journal} {Phys. Dark Univ.}\ }\textbf {\bibinfo
  {volume} {22}},\ \bibinfo {pages} {1} (\bibinfo {year} {2018})},\ \Eprint
  {http://arxiv.org/abs/1801.01509} {arXiv:1801.01509 [astro-ph.SR]}
  \BibitemShut {NoStop}%
\bibitem [{\citenamefont {Budker}\ \emph {et~al.}(2020)\citenamefont {Budker},
  \citenamefont {Flambaum}, \citenamefont {Liang},\ and\ \citenamefont
  {Zhitnitsky}}]{Budker:2019zka}%
  \BibitemOpen
  \bibfield  {author} {\bibinfo {author} {\bibfnamefont {D.}~\bibnamefont
  {Budker}}, \bibinfo {author} {\bibfnamefont {V.~V.}\ \bibnamefont
  {Flambaum}}, \bibinfo {author} {\bibfnamefont {X.}~\bibnamefont {Liang}}, \
  and\ \bibinfo {author} {\bibfnamefont {A.}~\bibnamefont {Zhitnitsky}},\
  }\href {\doibase 10.1103/PhysRevD.101.043012} {\bibfield  {journal} {\bibinfo
   {journal} {Phys. Rev. D}\ }\textbf {\bibinfo {volume} {101}},\ \bibinfo
  {pages} {043012} (\bibinfo {year} {2020})},\ \Eprint
  {http://arxiv.org/abs/1909.09475} {arXiv:1909.09475 [hep-ph]} \BibitemShut
  {NoStop}%
\bibitem [{\citenamefont {Liang}\ \emph {et~al.}(2021)\citenamefont {Liang},
  \citenamefont {Peshkov}, \citenamefont {Van~Waerbeke},\ and\ \citenamefont
  {Zhitnitsky}}]{Liang:2020mnz}%
  \BibitemOpen
  \bibfield  {author} {\bibinfo {author} {\bibfnamefont {X.}~\bibnamefont
  {Liang}}, \bibinfo {author} {\bibfnamefont {E.}~\bibnamefont {Peshkov}},
  \bibinfo {author} {\bibfnamefont {L.}~\bibnamefont {Van~Waerbeke}}, \ and\
  \bibinfo {author} {\bibfnamefont {A.}~\bibnamefont {Zhitnitsky}},\ }\href
  {\doibase 10.1103/PhysRevD.103.096001} {\bibfield  {journal} {\bibinfo
  {journal} {Phys. Rev. D}\ }\textbf {\bibinfo {volume} {103}},\ \bibinfo
  {pages} {096001} (\bibinfo {year} {2021})},\ \Eprint
  {http://arxiv.org/abs/2012.00765} {arXiv:2012.00765} \BibitemShut {NoStop}%
\bibitem [{\citenamefont {Zhang}\ \emph {et~al.}(2022)\citenamefont {Zhang},
  \citenamefont {Horns},\ and\ \citenamefont {Ghosh}}]{Zhang:2021bpa}%
  \BibitemOpen
  \bibfield  {author} {\bibinfo {author} {\bibfnamefont {Z.}~\bibnamefont
  {Zhang}}, \bibinfo {author} {\bibfnamefont {D.}~\bibnamefont {Horns}}, \ and\
  \bibinfo {author} {\bibfnamefont {O.}~\bibnamefont {Ghosh}},\ }\href
  {\doibase 10.1103/PhysRevD.106.023003} {\bibfield  {journal} {\bibinfo
  {journal} {Phys. Rev. D}\ }\textbf {\bibinfo {volume} {106}},\ \bibinfo
  {pages} {023003} (\bibinfo {year} {2022})},\ \Eprint
  {http://arxiv.org/abs/2111.04541} {arXiv:2111.04541 [hep-ex]} \BibitemShut
  {NoStop}%
\bibitem [{\citenamefont {Kahn}\ \emph {et~al.}(2016)\citenamefont {Kahn},
  \citenamefont {Safdi},\ and\ \citenamefont {Thaler}}]{Kahn:2016aff}%
  \BibitemOpen
  \bibfield  {author} {\bibinfo {author} {\bibfnamefont {Y.}~\bibnamefont
  {Kahn}}, \bibinfo {author} {\bibfnamefont {B.~R.}\ \bibnamefont {Safdi}}, \
  and\ \bibinfo {author} {\bibfnamefont {J.}~\bibnamefont {Thaler}},\ }\href
  {\doibase 10.1103/PhysRevLett.117.141801} {\bibfield  {journal} {\bibinfo
  {journal} {Phys. Rev. Lett.}\ }\textbf {\bibinfo {volume} {117}},\ \bibinfo
  {pages} {141801} (\bibinfo {year} {2016})},\ \Eprint
  {http://arxiv.org/abs/1602.01086} {arXiv:1602.01086 [hep-ph]} \BibitemShut
  {NoStop}%
\bibitem [{\citenamefont {Sikivie}\ \emph {et~al.}(2014)\citenamefont
  {Sikivie}, \citenamefont {Sullivan},\ and\ \citenamefont
  {Tanner}}]{Sikivie:2013laa}%
  \BibitemOpen
  \bibfield  {author} {\bibinfo {author} {\bibfnamefont {P.}~\bibnamefont
  {Sikivie}}, \bibinfo {author} {\bibfnamefont {N.}~\bibnamefont {Sullivan}}, \
  and\ \bibinfo {author} {\bibfnamefont {D.~B.}\ \bibnamefont {Tanner}},\
  }\href {\doibase 10.1103/PhysRevLett.112.131301} {\bibfield  {journal}
  {\bibinfo  {journal} {Phys. Rev. Lett.}\ }\textbf {\bibinfo {volume} {112}},\
  \bibinfo {pages} {131301} (\bibinfo {year} {2014})},\ \Eprint
  {http://arxiv.org/abs/1310.8545} {arXiv:1310.8545 [hep-ph]} \BibitemShut
  {NoStop}%
\bibitem [{\citenamefont {Chaudhuri}\ \emph {et~al.}(2018)\citenamefont
  {Chaudhuri}, \citenamefont {Irwin}, \citenamefont {Graham},\ and\
  \citenamefont {Mardon}}]{Chaudhuri:2018rqn}%
  \BibitemOpen
  \bibfield  {author} {\bibinfo {author} {\bibfnamefont {S.}~\bibnamefont
  {Chaudhuri}}, \bibinfo {author} {\bibfnamefont {K.}~\bibnamefont {Irwin}},
  \bibinfo {author} {\bibfnamefont {P.~W.}\ \bibnamefont {Graham}}, \ and\
  \bibinfo {author} {\bibfnamefont {J.}~\bibnamefont {Mardon}},\ }\href@noop {}
  {\  (\bibinfo {year} {2018})},\ \Eprint {http://arxiv.org/abs/1803.01627}
  {arXiv:1803.01627 [hep-ph]} \BibitemShut {NoStop}%
\bibitem [{\citenamefont {Gardikiotis}(2024)}]{Gardikiotis:2024crt}%
  \BibitemOpen
  \bibfield  {author} {\bibinfo {author} {\bibfnamefont {A.}~\bibnamefont
  {Gardikiotis}} (\bibinfo {collaboration} {MADMAX}),\ }\href {\doibase
  10.1002/andp.202300046} {\bibfield  {journal} {\bibinfo  {journal} {Annalen
  Phys.}\ }\textbf {\bibinfo {volume} {536}},\ \bibinfo {pages} {2300046}
  (\bibinfo {year} {2024})}\BibitemShut {NoStop}%
\bibitem [{\citenamefont {Bajjali}\ \emph {et~al.}(2023)\citenamefont {Bajjali}
  \emph {et~al.}}]{Bajjali:2023uis}%
  \BibitemOpen
  \bibfield  {author} {\bibinfo {author} {\bibfnamefont {F.}~\bibnamefont
  {Bajjali}} \emph {et~al.},\ }\href {\doibase 10.1088/1475-7516/2023/08/077}
  {\bibfield  {journal} {\bibinfo  {journal} {JCAP}\ }\textbf {\bibinfo
  {volume} {08}},\ \bibinfo {pages} {077} (\bibinfo {year} {2023})},\ \Eprint
  {http://arxiv.org/abs/2306.05934} {arXiv:2306.05934 [hep-ex]} \BibitemShut
  {NoStop}%
\bibitem [{\citenamefont {{Marsh}}(2016)}]{Marsh:2015xka}%
  \BibitemOpen
  \bibfield  {author} {\bibinfo {author} {\bibfnamefont {D.~J.~E.}\
  \bibnamefont {{Marsh}}},\ }\href {\doibase 10.1016/j.physrep.2016.06.005}
  {\bibfield  {journal} {\bibinfo  {journal} {Phys. Rep.}\ }\textbf {\bibinfo
  {volume} {643}},\ \bibinfo {pages} {1} (\bibinfo {year} {2016})},\ \Eprint
  {http://arxiv.org/abs/1510.07633} {arXiv:1510.07633} \BibitemShut {NoStop}%
\bibitem [{\citenamefont {{Graham}}\ \emph {et~al.}(2015)\citenamefont
  {{Graham}}, \citenamefont {{Irastorza}}, \citenamefont {{Lamoreaux}},
  \citenamefont {{Lindner}},\ and\ \citenamefont {{van
  Bibber}}}]{Graham:2015ouw}%
  \BibitemOpen
  \bibfield  {author} {\bibinfo {author} {\bibfnamefont {P.~W.}\ \bibnamefont
  {{Graham}}}, \bibinfo {author} {\bibfnamefont {I.~G.}\ \bibnamefont
  {{Irastorza}}}, \bibinfo {author} {\bibfnamefont {S.~K.}\ \bibnamefont
  {{Lamoreaux}}}, \bibinfo {author} {\bibfnamefont {A.}~\bibnamefont
  {{Lindner}}}, \ and\ \bibinfo {author} {\bibfnamefont {K.~A.}\ \bibnamefont
  {{van Bibber}}},\ }\href {\doibase 10.1146/annurev-nucl-102014-022120}
  {\bibfield  {journal} {\bibinfo  {journal} {Annual Review of Nuclear and
  Particle Science}\ }\textbf {\bibinfo {volume} {65}},\ \bibinfo {pages} {485}
  (\bibinfo {year} {2015})},\ \Eprint {http://arxiv.org/abs/1602.00039}
  {arXiv:1602.00039 [hep-ex]} \BibitemShut {NoStop}%
\bibitem [{\citenamefont {Battesti}\ \emph {et~al.}(2018)\citenamefont
  {Battesti} \emph {et~al.}}]{Battesti:2018bgc}%
  \BibitemOpen
  \bibfield  {author} {\bibinfo {author} {\bibfnamefont {R.}~\bibnamefont
  {Battesti}} \emph {et~al.},\ }\href {\doibase 10.1016/j.physrep.2018.07.005}
  {\bibfield  {journal} {\bibinfo  {journal} {Phys. Rept.}\ }\textbf {\bibinfo
  {volume} {765-766}},\ \bibinfo {pages} {1} (\bibinfo {year} {2018})},\
  \Eprint {http://arxiv.org/abs/1803.07547} {arXiv:1803.07547
  [physics.ins-det]} \BibitemShut {NoStop}%
\bibitem [{\citenamefont {Irastorza}\ and\ \citenamefont
  {Redondo}(2018)}]{Irastorza:2018dyq}%
  \BibitemOpen
  \bibfield  {author} {\bibinfo {author} {\bibfnamefont {I.~G.}\ \bibnamefont
  {Irastorza}}\ and\ \bibinfo {author} {\bibfnamefont {J.}~\bibnamefont
  {Redondo}},\ }\href {\doibase 10.1016/j.ppnp.2018.05.003} {\bibfield
  {journal} {\bibinfo  {journal} {Prog. Part. Nucl. Phys.}\ }\textbf {\bibinfo
  {volume} {102}},\ \bibinfo {pages} {89} (\bibinfo {year} {2018})},\ \Eprint
  {http://arxiv.org/abs/1801.08127} {arXiv:1801.08127 [hep-ph]} \BibitemShut
  {NoStop}%
\bibitem [{\citenamefont {Sikivie}(2021)}]{Sikivie:2020zpn}%
  \BibitemOpen
  \bibfield  {author} {\bibinfo {author} {\bibfnamefont {P.}~\bibnamefont
  {Sikivie}},\ }\href {\doibase 10.1103/RevModPhys.93.015004} {\bibfield
  {journal} {\bibinfo  {journal} {Rev. Mod. Phys.}\ }\textbf {\bibinfo {volume}
  {93}},\ \bibinfo {pages} {015004} (\bibinfo {year} {2021})},\ \Eprint
  {http://arxiv.org/abs/2003.02206} {arXiv:2003.02206 [hep-ph]} \BibitemShut
  {NoStop}%
\bibitem [{\citenamefont {Liang}\ \emph {et~al.}(2020)\citenamefont {Liang},
  \citenamefont {Mead}, \citenamefont {Siddiqui}, \citenamefont
  {Van~Waerbeke},\ and\ \citenamefont {Zhitnitsky}}]{Liang:2019lya}%
  \BibitemOpen
  \bibfield  {author} {\bibinfo {author} {\bibfnamefont {X.}~\bibnamefont
  {Liang}}, \bibinfo {author} {\bibfnamefont {A.}~\bibnamefont {Mead}},
  \bibinfo {author} {\bibfnamefont {M.~S.~R.}\ \bibnamefont {Siddiqui}},
  \bibinfo {author} {\bibfnamefont {L.}~\bibnamefont {Van~Waerbeke}}, \ and\
  \bibinfo {author} {\bibfnamefont {A.}~\bibnamefont {Zhitnitsky}},\ }\href
  {\doibase 10.1103/PhysRevD.101.043512} {\bibfield  {journal} {\bibinfo
  {journal} {Phys. Rev. D}\ }\textbf {\bibinfo {volume} {101}},\ \bibinfo
  {pages} {043512} (\bibinfo {year} {2020})},\ \Eprint
  {http://arxiv.org/abs/1908.04675} {arXiv:1908.04675 [astro-ph.CO]}
  \BibitemShut {NoStop}%
\bibitem [{\citenamefont {Salucci}\ \emph {et~al.}(2020)\citenamefont
  {Salucci}, \citenamefont {Turini},\ and\ \citenamefont
  {Di~Paolo}}]{Salucci:2020eqo}%
  \BibitemOpen
  \bibfield  {author} {\bibinfo {author} {\bibfnamefont {P.}~\bibnamefont
  {Salucci}}, \bibinfo {author} {\bibfnamefont {N.}~\bibnamefont {Turini}}, \
  and\ \bibinfo {author} {\bibfnamefont {C.}~\bibnamefont {Di~Paolo}},\ }\href
  {\doibase 10.1142/9789811233913_0075} {\bibfield  {journal} {\bibinfo
  {journal} {Universe}\ }\textbf {\bibinfo {volume} {6}},\ \bibinfo {pages}
  {118} (\bibinfo {year} {2020})},\ \Eprint {http://arxiv.org/abs/2008.04052}
  {arXiv:2008.04052 [astro-ph.CO]} \BibitemShut {NoStop}%
\bibitem [{\citenamefont {Zhitnitsky}(2023)}]{Zhitnitsky:2023znn}%
  \BibitemOpen
  \bibfield  {author} {\bibinfo {author} {\bibfnamefont {A.}~\bibnamefont
  {Zhitnitsky}},\ }\href {\doibase 10.1016/j.dark.2023.101217} {\bibfield
  {journal} {\bibinfo  {journal} {Phys. Dark Univ.}\ }\textbf {\bibinfo
  {volume} {40}},\ \bibinfo {pages} {101217} (\bibinfo {year} {2023})},\
  \Eprint {http://arxiv.org/abs/2302.00010} {arXiv:2302.00010 [hep-ph]}
  \BibitemShut {NoStop}%
\bibitem [{\citenamefont {Henry}\ \emph {et~al.}(2014)\citenamefont {Henry},
  \citenamefont {Murthy}, \citenamefont {Overduin},\ and\ \citenamefont
  {Tyler}}]{Henry_2014}%
  \BibitemOpen
  \bibfield  {author} {\bibinfo {author} {\bibfnamefont {R.~C.}\ \bibnamefont
  {Henry}}, \bibinfo {author} {\bibfnamefont {J.}~\bibnamefont {Murthy}},
  \bibinfo {author} {\bibfnamefont {J.}~\bibnamefont {Overduin}}, \ and\
  \bibinfo {author} {\bibfnamefont {J.}~\bibnamefont {Tyler}},\ }\href
  {\doibase 10.1088/0004-637x/798/1/14} {\bibfield  {journal} {\bibinfo
  {journal} {The Astrophysical Journal}\ }\textbf {\bibinfo {volume} {798}},\
  \bibinfo {pages} {14} (\bibinfo {year} {2014})}\BibitemShut {NoStop}%
\bibitem [{\citenamefont {Akshaya}\ \emph {et~al.}(2018)\citenamefont
  {Akshaya}, \citenamefont {Murthy}, \citenamefont {Ravichandran},
  \citenamefont {Henry},\ and\ \citenamefont {Overduin}}]{Akshaya_2018}%
  \BibitemOpen
  \bibfield  {author} {\bibinfo {author} {\bibfnamefont {M.~S.}\ \bibnamefont
  {Akshaya}}, \bibinfo {author} {\bibfnamefont {J.}~\bibnamefont {Murthy}},
  \bibinfo {author} {\bibfnamefont {S.}~\bibnamefont {Ravichandran}}, \bibinfo
  {author} {\bibfnamefont {R.~C.}\ \bibnamefont {Henry}}, \ and\ \bibinfo
  {author} {\bibfnamefont {J.}~\bibnamefont {Overduin}},\ }\href {\doibase
  10.3847/1538-4357/aabcb9} {\bibfield  {journal} {\bibinfo  {journal} {The
  Astrophysical Journal}\ }\textbf {\bibinfo {volume} {858}},\ \bibinfo {pages}
  {101} (\bibinfo {year} {2018})}\BibitemShut {NoStop}%
\bibitem [{\citenamefont {{Akshaya}}\ \emph {et~al.}(2019)\citenamefont
  {{Akshaya}}, \citenamefont {{Murthy}}, \citenamefont {{Ravichandran}},
  \citenamefont {{Henry}},\ and\ \citenamefont
  {{Overduin}}}]{2019MNRAS.489.1120A}%
  \BibitemOpen
  \bibfield  {author} {\bibinfo {author} {\bibfnamefont {M.~S.}\ \bibnamefont
  {{Akshaya}}}, \bibinfo {author} {\bibfnamefont {J.}~\bibnamefont {{Murthy}}},
  \bibinfo {author} {\bibfnamefont {S.}~\bibnamefont {{Ravichandran}}},
  \bibinfo {author} {\bibfnamefont {R.~C.}\ \bibnamefont {{Henry}}}, \ and\
  \bibinfo {author} {\bibfnamefont {J.}~\bibnamefont {{Overduin}}},\ }\href
  {\doibase 10.1093/mnras/stz2186} {\bibfield  {journal} {\bibinfo  {journal}
  {Mon.\ Not.\ R.\ Astron.\ Soc.}\ }\textbf {\bibinfo {volume} {489}},\
  \bibinfo {pages} {1120} (\bibinfo {year} {2019})},\ \Eprint
  {http://arxiv.org/abs/1908.02260} {arXiv:1908.02260 [astro-ph.GA]}
  \BibitemShut {NoStop}%
\bibitem [{\citenamefont
  {Zhitnitsky}(2022{\natexlab{b}})}]{Zhitnitsky:2021wjb}%
  \BibitemOpen
  \bibfield  {author} {\bibinfo {author} {\bibfnamefont {A.}~\bibnamefont
  {Zhitnitsky}},\ }\href {\doibase 10.1016/j.physletb.2022.137015} {\bibfield
  {journal} {\bibinfo  {journal} {Phys. Lett. B}\ }\textbf {\bibinfo {volume}
  {828}},\ \bibinfo {pages} {137015} (\bibinfo {year} {2022}{\natexlab{b}})},\
  \Eprint {http://arxiv.org/abs/2110.05489} {arXiv:2110.05489 [hep-ph]}
  \BibitemShut {NoStop}%
\bibitem [{\citenamefont {{Zhitnitsky}}(2017)}]{Zhitnitsky:2017rop}%
  \BibitemOpen
  \bibfield  {author} {\bibinfo {author} {\bibfnamefont {A.}~\bibnamefont
  {{Zhitnitsky}}},\ }\href {\doibase 10.1088/1475-7516/2017/10/050} {\bibfield
  {journal} {\bibinfo  {journal} {\jcap}\ }\textbf {\bibinfo {volume} {10}},\
  \bibinfo {eid} {050} (\bibinfo {year} {2017})},\ \Eprint
  {http://arxiv.org/abs/1707.03400} {arXiv:1707.03400 [astro-ph.SR]}
  \BibitemShut {NoStop}%
\bibitem [{\citenamefont {Raza}\ \emph {et~al.}(2018)\citenamefont {Raza},
  \citenamefont {van Waerbeke},\ and\ \citenamefont
  {Zhitnitsky}}]{Raza:2018gpb}%
  \BibitemOpen
  \bibfield  {author} {\bibinfo {author} {\bibfnamefont {N.}~\bibnamefont
  {Raza}}, \bibinfo {author} {\bibfnamefont {L.}~\bibnamefont {van Waerbeke}},
  \ and\ \bibinfo {author} {\bibfnamefont {A.}~\bibnamefont {Zhitnitsky}},\
  }\href {\doibase 10.1103/PhysRevD.98.103527} {\bibfield  {journal} {\bibinfo
  {journal} {Phys. Rev. D}\ }\textbf {\bibinfo {volume} {98}},\ \bibinfo
  {pages} {103527} (\bibinfo {year} {2018})},\ \Eprint
  {http://arxiv.org/abs/1805.01897} {arXiv:1805.01897 [astro-ph.SR]}
  \BibitemShut {NoStop}%
\bibitem [{\citenamefont {{Parker}}(1988)}]{Parker}%
  \BibitemOpen
  \bibfield  {author} {\bibinfo {author} {\bibfnamefont {E.~N.}\ \bibnamefont
  {{Parker}}},\ }\href {\doibase 10.1086/166485} {\bibfield  {journal}
  {\bibinfo  {journal} {\apj}\ }\textbf {\bibinfo {volume} {330}},\ \bibinfo
  {pages} {474} (\bibinfo {year} {1988})}\BibitemShut {NoStop}%
\bibitem [{\citenamefont {Homola}\ \emph {et~al.}(2020)\citenamefont {Homola}
  \emph {et~al.}}]{sym12111835}%
  \BibitemOpen
  \bibfield  {author} {\bibinfo {author} {\bibfnamefont {P.}~\bibnamefont
  {Homola}} \emph {et~al.},\ }\href {\doibase 10.3390/sym12111835} {\bibfield
  {journal} {\bibinfo  {journal} {Symmetry}\ }\textbf {\bibinfo {volume} {12}}
  (\bibinfo {year} {2020}),\ 10.3390/sym12111835}\BibitemShut {NoStop}%
\end{thebibliography}%

\end{document}